\documentclass[apj]{aastex631}
\usepackage{amsmath}

\shorttitle{Dark Matter in Fractional Gravity}
\shortauthors{F. Benetti et al.}

\begin{document}

\title{Dark Matter in Fractional Gravity I: Astrophysical Tests on Galactic Scales} 

\author[0000-0002-2778-9131]{Francesco Benetti}\affiliation{SISSA, Via Bonomea 265, 34136 Trieste, Italy}

\author[0000-0002-4882-1735]{Andrea Lapi}
\affiliation{SISSA, Via Bonomea 265, 34136 Trieste, Italy}\affiliation{IFPU - Institute for fundamental physics of the Universe, Via Beirut 2, 34014 Trieste, Italy}\affiliation{INFN-Sezione di Trieste, via Valerio 2, 34127 Trieste,  Italy}\affiliation{IRA-INAF, Via Gobetti 101, 40129 Bologna, Italy}

\author[0000-0003-3248-5666]{Giovanni Gandolfi}\affiliation{SISSA, Via Bonomea 265, 34136 Trieste, Italy}\affiliation{IFPU - Institute for fundamental physics of the Universe, Via Beirut 2, 34014 Trieste, Italy}\affiliation{INFN-Sezione di Trieste, via Valerio 2, 34127 Trieste, Italy}

\author[0000-0002-5476-2954]{Paolo Salucci}\affiliation{SISSA, Via Bonomea 265, 34136 Trieste, Italy}\affiliation{IFPU - Institute for fundamental physics of the Universe, Via Beirut 2, 34014 Trieste, Italy}\affiliation{INFN-Sezione di Trieste, via Valerio 2, 34127 Trieste, Italy}

\author[0000-0003-1186-8430]{Luigi Danese}\affiliation{SISSA, Via Bonomea 265, 34136 Trieste, Italy}\affiliation{IFPU - Institute for fundamental physics of the Universe, Via Beirut 2, 34014 Trieste, Italy}

\begin{abstract}
We explore the possibility that the dark matter (DM) component in galaxies may originate fractional gravity. In such a framework, the standard law of inertia continues to hold, but the gravitational potential associated to a given DM density distribution is determined by a modified Poisson equation including fractional derivatives (i.e., derivatives of non-integer type), that are meant to describe non-local effects. We derive analytically the expression of the potential that in fractional gravity corresponds to various spherically symmetric density profiles, including the Navarro-Frenk-White (NFW) distribution that is usually exploited to describe virialized halos of collisionless DM as extracted from $N-$body cosmological simulations. We show that in fractional gravity the dynamics of a test particle moving in a cuspy NFW density distribution is substantially altered with respect to the Newtonian case (i.e., basing on the standard Poisson equation), mirroring what in Newtonian gravity would instead be sourced by a density profile with an inner core. We test the fractional gravity framework on galactic scales, showing that:
(i) it can provide accurate fits to the stacked rotation curves of spiral galaxies with different properties, including dwarfs and high/low surface-brightness systems; (ii) it can reproduce to reasonable accuracy the observed shape and scatter of the radial acceleration relation (RAR) over an extended range of galaxy accelerations; (iii) it can properly account for the universal surface density and the core radius vs. disk scale-length scaling relations as empirically measured in different galaxies over an extended mass range. Interestingly, the analysis of galaxy RCs also suggests that the strength of fractional gravity effects gets progressively weaker in more massive systems; therefore our fractional-gravity framework can substantially alleviate the small-scale issues of the standard DM paradigm in Newtonian gravity, while saving its successes on large cosmological scales. Finally, we discuss the possible origin of the fractional gravity behavior as a fundamental or emerging property of the elusive DM component.
\end{abstract}

\keywords{Cosmology (343) - Dark matter (353) - Non-standard theories of gravity (1118)}

\section{Introduction} \label{sec|intro}

A multitude of astrophysical and cosmological probes have firmly established that baryons constitute only some $15\%$ of the total matter content in the Universe, the rest being in the form of a `dark matter’ (DM) component. A non-exhaustive list of such evidences include: kinematics of spiral galaxies (e.g., Rubin et al. 1980; Persic et al. 1996; see review by Salucci 2019 and references therein); dynamical modeling of ellipticals (e.g., Bacon et al. 2001; Cappellari et al. 2013; Mendel et al. 2020); cosmic microwave background and big-bang nucleosynthesis constraints (e.g., Bennett et al. 2003; Ade et al. 2014; Aver et al. 2015: Aghanim et al. 2020); baryon acoustic oscillations measurements (e.g., Eisenstein et al. 2005; Beutler et al. 2011; Zhao et al. 2022); type-I$a$ cosmography (e.g., Perlmutter et al. 1999; Scolnic et al. 2018; Brout et al. 2022); cosmic shear galaxy surveys (e.g., Heymans et al. 2013; Amon et al. 2022; Secco et al. 2022); X-ray, Sunyaev-Zel'dovich, strong/weak lensing observations and number counts of galaxy clusters (e.g., White et al. 1993; Allen et al. 2011; Umetsu et al. 2020; Garrell et al. 2022; Mantz et al. 2022); observations of the 'Bullet Cluster' (e.g., Markevitch et al. 2004; Clowe et al. 2006; Paraficz et al. 2016). Incidentally, note that most of the baryons in the Universe are locked in a hot/warm intergalactic medium, with only a small fraction (less than about $20\%$) residing as cold gas and stars in galaxies (e.g., Tumlinson et al. 2017 and references therein).

Still, no firm detection of DM particles has been made so far, despite the big efforts carried on with colliders (see Mitsou et al. 2013; Argyropoulos et al. 2021) or with direct (see Aprile et al. 2018; Bernabei et al. 2020) and indirect (see Ackermann et al. 2017; Zornoza et al. 2021) searches in the sky. The standard picture envisages DM to be constituted by weakly interacting particles with GeV masses (see Bertone \& Hooper 2018), that are non-relativistic at the epoch of decoupling (hence they are dubbed 'cold' dark matter or CDM) and feature negligible free-streaming velocities (i.e., they do not diffuse out of perturbations before collapse). As a consequence, bound CDM structures called halos grow sequentially in time and hierarchically in mass by stochastically merging together (e.g., Frenk \& White 2012; Lapi et al. 2020, 2022). 

Although on cosmological scales the CDM hypothesis is remarkably consistent with the data, on (sub)galactic scales it faces some severe challenges. For example, with respect to the predictions of gravity-only $N$-body simulations, the shape of the inner density profiles inferred from the rotation curve (RC) in DM-dominated dwarfs is too flat (e.g., Flores \& Primack 1994; Gentile et al. 2004; de Blok et al. 2008; Oh et al. 2015; Salucci et al. 2021), and the number and dynamical properties of observed Milky Way satellites differ from those of subhalos (see Boylan-Kolchin et al. 2012; Bullock \& Boylan-Kolchin 2017). Moreover, the emergence of tight empirical relationships between properties of the dark and luminous components in disk-dominated galaxies, such as the radial acceleration relation (RAR; see McGaugh et al. 2016; Lelli et al. 2017; also Di Paolo et al. 2019), the universal core surface density (see Donato et al. 2009) and the scaling between the core radius and the disk scale-length (see Donato et al. 2004) are puzzling and seem to be indicative of a new dark sector and/or of non-gravitational coupling between DM particles and baryons (see Salucci et al. 2020). Some of the above findings can in principle be explained in CDM by invoking physical processes causing transfer of energy and angular momentum from the baryons to DM particles, such as dynamical friction (see El-Zant et al. 2001; Tonini et al. 2006) or feedback effects from stars and active galactic nuclei (see Pontzen \& Governato 2014; Freundlich et al. 2020), but in recent years it has become progressively clearer that a fine-tuning for these solutions is required to explain in detail the current observations.

This has triggered the consideration of alternative, and perhaps more fascinating, explanations that rely on non-standard particle candidates 
(see Bertone et al. 2004; Adhikari et al. 2017; Salucci et al. 2021); the most widespread possibilities include keV-scale warm DM particles (see Bode et al. 2001; Lovell et al. 2014), fuzzy or particle-wave DM (see Hu et al. 2000; Hui et al. 2017), self-interacting dark matter (see Vogelsberger et al. 2016), dark photon-dark matter (McDermott \& Witte 2020; Bolton et al. 2022). As a consequence of free-streaming, quantum pressure effects, and/or dark-sector interactions, all these scenarios produce a matter power spectrum suppressed on small scales, fewer (sub)structures, and flatter inner density profiles within halos relative to CDM. However, all these possibilities, despite being still viable to some extent, have been strongly constrained by various astrophysical probes, most noticeably Lyman-$\alpha$ forest (e.g., Irsic et al. 2017), high-redshift galaxy counts (e.g., Shirasaki et al. 2021), cosmic reionization (e.g., Lapi et al. 2022), gravitational lensing (e.g., Ritondale et al. 2018), integrated 21 cm data (e.g., Carucci et al. 2015), $\gamma$-ray emission (e.g., Grand \& White 2022), fossil records of the Local Group (e.g., Weisz et al. 2017), dwarf galaxy profiles and scaling relations (e.g., Burkert 2020), Milky Way satellite galaxies (e.g., Newton et al. 2021), and cosmic star formation rate density (e.g., Gandolfi et al. 2022a), or a combination of these.

All that has motivated, on different grounds, the emergence of modified gravity theories. Perhaps the most famous one is constituted by the modified Newtonian dynamics (MOND) that was originally proposed by Milgrom (1983) and
further investigated in a rich literature (see Bekenstein 2004; Bruneton \& Esposito-Farese 2007; Famaey \& McGaugh 2012). As the name suggests, MOND aims to
explain galactic dynamics through a modification
of Newtonian gravity (or more generally of the Newton
second law) that comes into action at accelerations well below
a definite universal threshold; in its original formulation, DM is
not included and baryons are the only source of the
gravitational field. In fact, it
has been claimed that MOND (or theories reducing to it in the
weak-field limit) can properly fit galactic RCs (de Blok \&
McGaugh 1998; Sanders \& McGaugh 2002), and provide a
satisfying description of the RAR (e.g., Li et al. 2018).
On the other hand, the performances of MOND are highly debated when moving to galaxy clusters and to cosmological scales (see Aguirre et al. 2001; Scott et al. 2001; Angus et al. 2006; McGaugh 2015; Nieuwenhuizen 2017; Boran et al. 2018). Other more recent proposals include, e.g., an emergent (entropic) theory  of gravity (see Verlinde 2017; Yoon et al. 2023) or a dynamical non-minimal coupling between the DM component and the gravitational metric (see Gandolfi et al. 2021, 2022b).

In this paper we explore yet another possibility that has some features in common with modified gravity models, though being fundamentally different. Specifically, we propose that DM in galaxies originates fractional gravity; this means that the gravitational potential associated to a given DM density distribution is determined by a modified Poisson equation that includes fractional derivatives (i.e., derivatives of non-integer type), meant to describe non-local effects. Fractional calculus has found applications in the descriptions of several physical phenomena that subtends non-local behavior in space and time (e.g. material science, rheology, seismology, fluid dynamics, nuclear physics, medicine, finance, etc.). In astrophysics and cosmology the related investigations are still scanty, though their number and variety is progressively increasing in recent years (e.g., Calcagni 2010, 2013, 2021; Calcagni \& Varieschi 2022; García-Aspeitia et al. 2022; Giusti 2020; Giusti et al. 2020; Varieschi 2020, 2021, 2022, 2023; see also Sect. \ref{sec|MOND}).

Here in particular we aim at testing whether fractional gravity as originated by the DM component can solve the aforementioned problems of the standard paradigm in galaxies, while retaining its successes on cosmological scales. We stress that in our fractional gravity framework the law of inertia is unchanged, while only the Poisson equation is modified. Note that fractional gravity is not necessarily meant to be an ab-initio theory, but may constitute an effective description for a whole class of models that imply the development of non-local effects in the gravitational behavior of the DM component. As for the baryons, we assume that they originate standard gravity, though feeling the overall gravitational potential of the system.

We will show that in fractional gravity the dynamics of a test particle in the potential sourced by a standard NFW density distribution (that we consider as a representative rendition of the density distribution for halos of non-interacting DM) is substantially altered with respect to the Newtonian case. Specifically, such a framework can: (i) provide accurate fits to the stacked RCs of spiral galaxies with different properties, including dwarfs and high/low surface-brightness systems; (ii) reproduce to reasonable accuracy the observed shape and scatter of the RAR over an extended range of galaxy accelerations; (iii) properly account for the universal surface density and for the scaling relation between the core radius of the DM component and the disk scale-length.

The plan of the paper is as follows. In Section \ref{sec|methods} we introduce the fractional gravity framework and describe how this can be applied to fit the stacked RCs of spiral galaxies and infer related physical parameters via a Bayesian MCMC framework. In Section \ref{sec|results} we present and discuss our results, with particular focus on the RC of dwarf galaxies, on the halo to (stellar) disk mass relation, on the RAR and on the universal core surface density. In Sect. \ref{sec|MOND} we explore an alternative MONDian viewpoint to fractional gravity without DM.
In Section \ref{sec|summary} we summarize our findings and outline future perspectives and applications of the fractional gravity framework. Appendix \ref{app|fracalc} contains a primer on fractional calculus, including the fractional Laplacian and Poisson equation. In Appendix \ref{app|solutions} we provide original analytic solutions of the gravitational potential in fractional gravity for various spherically symmetric density distributions.

Throughout this work, we adopt the standard flat $\Lambda$CDM cosmology (see Aghanim et al. 2020) with rounded parameter values: matter density $\Omega_M = 0.3$, dark
energy density $\Omega_\Lambda = 0.7$, baryon density $\Omega_b = 0.05$, and
Hubble constant $H_0 = 100\,h$ km s$^{-1}$ Mpc$^{-1}$ with $h = 0.7$. Unless
otherwise specified, $G\approx 6.67\times 10^{-8}$ cm$^3$ g$^{-1}$ s$^{-2}$ indicates the standard gravitational constant.

\section{Methods and Analysis} \label{sec|methods}

As anticipated in Section \ref{sec|intro}, in this paper we propose that the kinematics observed in galactic structures may be explained by assuming that the DM component originates fractional gravity. In such a framework, although the standard law of inertia continues to hold, the gravitational potential associated to a given DM distribution is determined by a modified Poisson equation of the fractional type. The latter has inherently a non-local nature, and can profoundly alter the dynamics of a test particle with respect to the Newtonian case (i.e., basing on the standard Poisson equation).  

In this section we first present our new solution for the potential sourced by the NFW density profile in fractional gravity, and then describe how this can be applied to analyze stacked RCs of spiral galaxies via a Bayesian MCMC technique, so inferring relevant galaxy properties and at the same time constraining the parameters ruling the strength and length-scale of fractional gravity.

\subsection{Dark Matter in fractional gravity}

The density distribution of virialized halos for collisionless DM is routinely described via the Navarro-Frenk-White profile (see Navarro et al. 1997):
\begin{equation}
\rho(r) = \frac{\rho_s\,r_s^3}{r\,(r+r_s)^2}~;
\end{equation}
here $r_s$ is a scale radius and $\rho_s$ a characteristic density. The logarithmic density slope $\gamma\equiv {\rm d}\ln\rho/{\rm d}\ln r$ takes on values around $-3$ in the outskirts, $-2$ at $r\approx r_s$ and $-1$ in the inner region. The associated cumulative mass is given by
\begin{equation}
M(<r)=4\pi\,\int_0^r{\rm d}r'\,r'^2\,\rho(r')=M_s\, \left[\ln\left(1+\frac{r}{r_s}\right)-\frac{r/r_s}{1+r/r_s}\right]~,
\end{equation}
where $M_s\equiv 4\pi\,\rho_s\,r_s^3$.

In Newtonian gravity, the potential $\Phi(r)$ associated to a given density distribution $\rho(r)$ is computed from the Poisson equation supplemented with appropriate boundary conditions (usually taken as a vanishing potential at infinity):
\begin{equation}
\Delta\Phi(\mathbf{r})=4\pi G\, \rho(\mathbf{r})
\end{equation}
where $\Delta$ is the Laplacian operator. For the spherically symmetric NFW profile it is easily found that
\begin{equation}
\Phi(r) = -\frac{G M_s}{r}\,\,\log\left(1+\frac{r}{r_s}\right)~.
\end{equation}
The rotational velocity, or in other words the RC, $v^2(r) = r\,|{\rm d}\Phi/{\rm d}r|$ of a test mass in such a potential is just 
\begin{equation}
v^2(r) = \frac{G M_s}{r_s} \left[\frac{\ln(1+r/r_s)}{r/r_s}-\frac{1}{1+r/r_s}\right]~;
\end{equation}
this also equals $v^2(r)=G\, M(<r)/r$ as a consequence of Birkhoff theorem.

\begin{figure}[!t]
\centering
\includegraphics[width=0.6\textwidth]{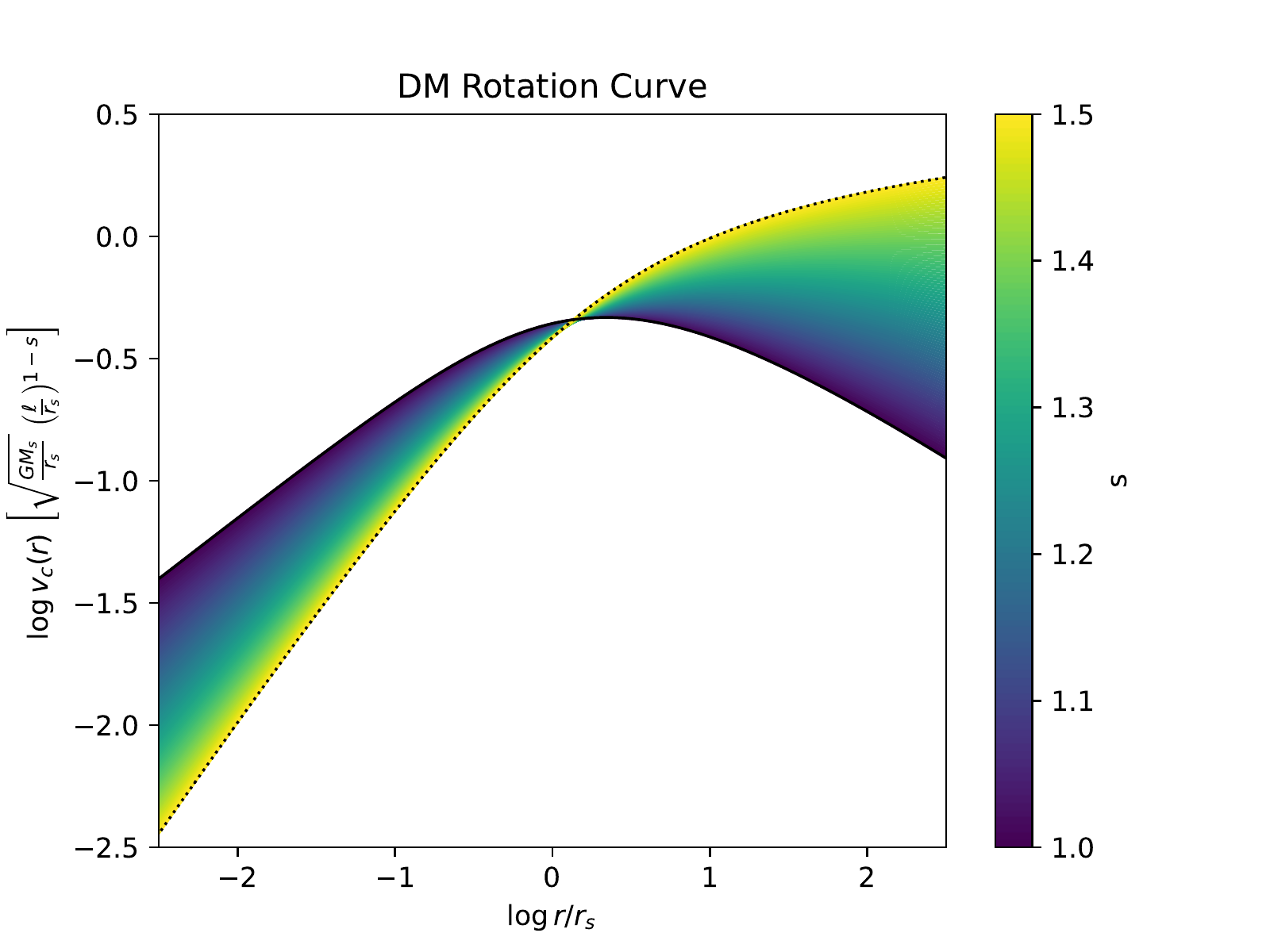}
\includegraphics[width=0.6\textwidth]{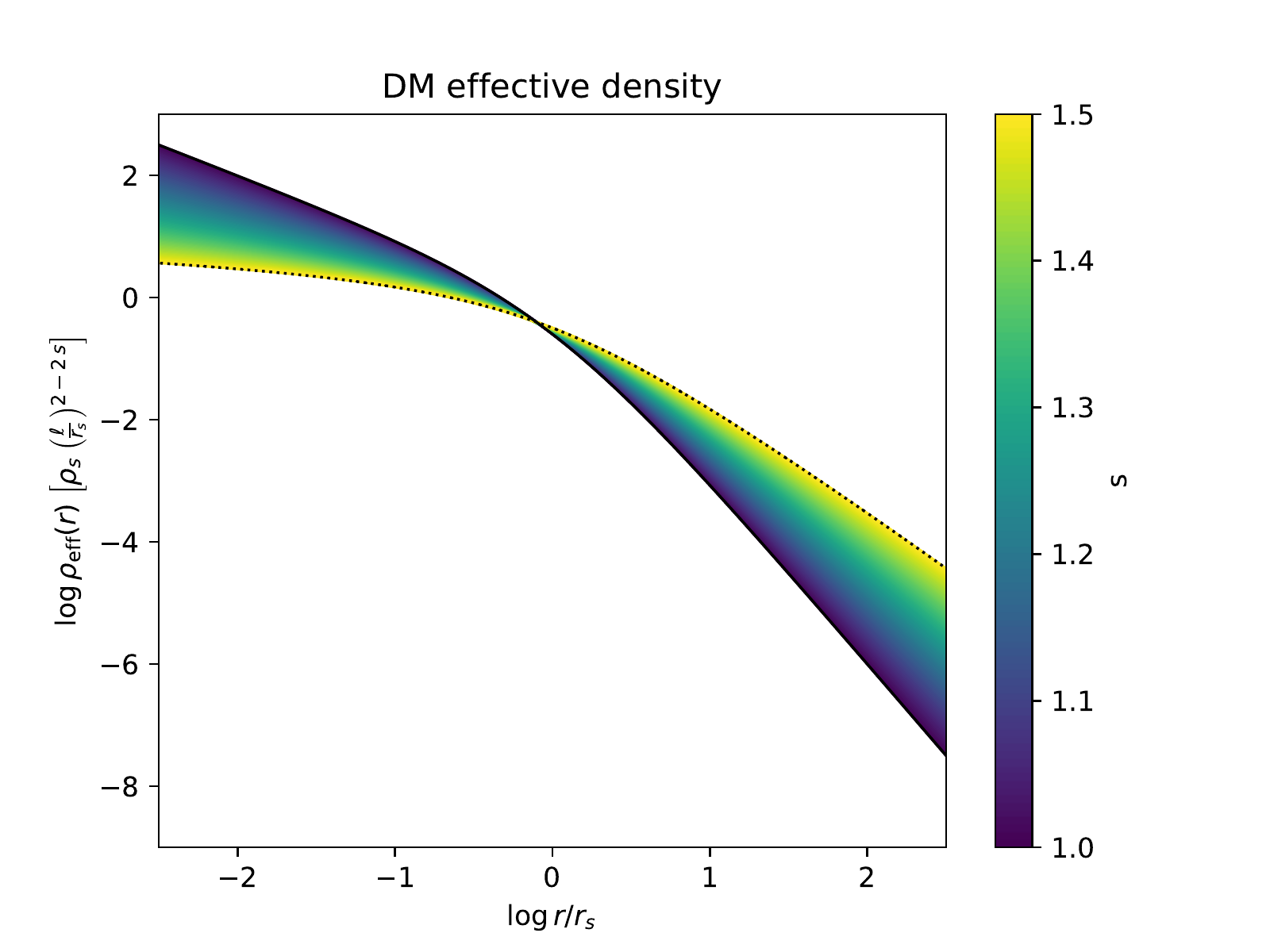}
\caption{DM rotation curve (top) and effective density (bottom) for different values of the fractional parameter $s$ (color-coded). For reference, the dotted line refers to the maximal value $s=3/2$.}\label{fig|model}
\end{figure}

In fractional gravity, the potential is instead derived from the modified Poisson equation (see Giusti 2020; more details can be found in Appendix \ref{app|fracalc})
\begin{equation}\label{eq|fracPoi}
(-\Delta)^s\, \Phi (\mathbf{r}) = -4\pi G\, \ell^{2-2s}\,\rho(\mathbf{r})
\end{equation}
where $(-\Delta)^s$ is the fractional Laplacian, $s\in [1,3/2]$ is the fractional index (this range of values for $s$ is required to avoid divergences, see Appendix \ref{app|fracalc}), and $\ell$ is a fractional length-scale (more on the meaning of these parameters at the end of this Section). We have solved the fractional Poisson equation sourced by the NFW density distribution, as detailed in Appendix \ref{app|solutions}. For $s\in [1,3/2)$ we find the solution
\begin{equation}\label{eq|pots}
\begin{aligned}
\Phi_{s}(r) = & -\frac{G M_s}{r_s}\,\frac{1}{2^{2s}\,\sqrt{\pi}}\,\left(\frac{\ell}{r_s}\right)^{2-2s}\,\frac{\Gamma\left(\frac{3}{2}-s\right)}{\Gamma(s+1)}\,\frac{r_s}{r}\,\left\{\frac{2\pi s}{\sin(2\pi s)}\,\left[\left(1+\frac{r}{r_s}\right)^{2s-2} - \left(1-\frac{r}{r_s}\right)^{2s-2}\right]+\right.\\
& \\
& +\left. \frac{(r/r_s)^{2s}}{1-(r/r_s)^{2}}\,\left[\left(1+\frac{r}{r_s}\right)\, _{2}F_{1}\left(1,1,2s+1,\frac{r}{r_s}\right) + \left(1-\frac{r}{r_s}\right)\, _{2} F_{1}\left(1,1,2s+1,-\frac{r}{r_s}\right)-\frac{4s}{2s-1} \right] \right\}~,
\end{aligned}
\end{equation}
in terms of the Euler Gamma function $\Gamma$ and of the ordinary hypergeometric function  ${}_2F_1$; it is straightforward to verify that $\Phi_{s=1}(r)$ coincides with the Newtonian expression. For the limiting case $s=3/2$ we instead find the solution
\begin{equation}\label{eq|potslim}
\begin{aligned}
\Phi_{s=3/2}(r) =& -\frac{G\,M_s}{\ell}\,\frac{1}{\pi}\,\frac{r_s}{r}\,	
\left\{2\,\frac{r}{r_s}\, \left[\log \left(\frac{r}{r_s}\right)-1\right]-\left(1+\frac{r}{r_s}\right)\,\log \left(\frac{r}{r_s}\right)\,\log \left(1+\frac{r}{r_s}\right) + \right.\\
&\\
&  +\left. \left(\frac{r}{r_s}-1\right)\, \text{Li}_2\left(1-\frac{r}{r_s}\right)-\left(1+\frac{r}{r_s}\right)\, \text{Li}_2\left(-\frac{r}{r_s}\right) + \frac{\pi^{2}}{6}\right\}~,
\end{aligned}
\end{equation}
in terms of the dilogarithm function ${\rm Li}_2$. The corresponding RC computed after $v_s^2(r) = r\,|{\rm d}\Phi_s/{\rm d}r|$, which remains true since the standard law of inertia continues to hold in fractional gravity, is illustrated in Fig. \ref{fig|model} (top panel) for different values of the fractional index $s$. For $s$ increasing above one (corresponding to Newtonian gravity), we find that the RC is steeper in the inner region, behaving as $v_s(r)\propto r^{s-1/2}$ for $s<3/2$ and $v_{s=3/2}(r)\propto r\,\sqrt{-\log r/r_s}$ for $s=3/2$; curiously, for the limiting value $s=3/2$ the RC tends to closely mirror a (Newtonian) solid body. Contrariwise, in the outskirts the RC in fractional gravity is appreciably flatter than the Newtonian case, behaving as $v_s(r)\propto r^{s-3/2}$ for $s<3/2$ and as $v_{s=3/2}(r)\propto \sqrt{\log r/r_s}$ for $s=3/2$.

In fractional gravity the Birkhoff theorem does not hold, but one can insist in writing $v_s^2(r) = G\, M_{{\rm eff},s}(<r)/r$ in terms of an effective mass $M_{{\rm eff}, s}(<r)$, and then differentiate this expression to obtain an effective density profile via $\rho_{{\rm eff},s}(r) = (1/4\pi\,r^2)\times {\rm d}M_{{\rm eff},s}/{\rm d}r$. This is actually the density behavior that one would infer by looking at the RC and interpreting the result in terms of Newtonian gravity. We illustrate such an effective density profile in Fig. \ref{fig|model} (bottom panel) for different values of the fractional index $s$. With $s$ increasing from unity (Newtonian case) the effective density profile progressively flattens. Specifically, in the inner region a cored-like behavior tends to be enforced, while in the outskirts the effective profile resembles an isothermal sphere. To have a grasp on the overall effect, consider the $s=3/2$ case where the effective density has a particularly simple analytic expression, given by
\begin{equation}
\rho_{{\rm eff}, s=3/2}(r) =\frac{2\, \rho_s\,r_s}{\pi\, \ell}\, \frac{\ln(r/r_s)}{(r/r_s)^2-1}~;
\end{equation}
interestingly, the logarithmic slope of this profile is $-2$ in the outskirts, $-1$ at $r\approx r_s$ and $\lesssim -0.5$ for $r\lesssim r_s/10$. All in all, the effective density is practically indistinguishable from the run of a cored pseudo-isothermal sphere. This slope around $r\sim 0.1\, r_s$ is remarkably close to the observational value estimated from the kinematic of individual dwarf galaxies (e.g., Oh et al. 2015; Salucci et al. 2021).

The physical reason behind this behavior is that the fractional Laplacian is inherently non-local; the effect is markedly evident in Fourier space, where it is easily found that $\mathcal{F}\{\rho_{{\rm eff},s}\}/\mathcal{F}\{\rho\}\propto (\lambda/\ell)^{2s-2}$ in terms of the Fourier transform $\mathcal{F}$ and of the wavelength $\lambda=2\pi/k$ associated to the Fourier mode with wavenumber $k$. This means that fractional gravity causes, with respect to the Newtonian case, a transfer of power from modes with $\lambda\lesssim \ell$, which are suppressed, to modes with $\lambda\gtrsim \ell$, which are instead enhanced. Therefore the parameter $s$ measures the strength of the non-locality, while the length-scale $\ell$ can be interpreted as the typical size 
below which gravitational effects are somewhat reduced and above which they are instead amplified by non-locality (around $r\approx \ell$ the dynamics is almost unaffected and indistinguishable from the Newtonian case).

\subsection{Bayesian analysis of stacked RCs}\label{sec|Bayes}

We test the fractional gravity framework by fitting observed stacked RCs of local spiral galaxies; these include high surface brightness (HSB; Persic et al. 1996; Yegorova \& Salucci 2007) systems divided in $11$ optical radial/velocity bins\footnote{The optical radius is $r_{\rm opt}\approx 3.2\,r_{\rm d}$ in terms of the exponential scale-length $r_{\rm d}$ of the (stellar) disk; the optical velocity is the circular velocity at the optical radius.}, low surface brightness (LSB; see Di Paolo et al. 2019; Dehghani et al. 2020) systems divided in $5$ bins, and $1$ bin for dwarf galaxies (Dw; see Karukes \& Salucci 2017). The stacked RCs are built by co-adding high-quality, high-resolution individual data of thousands galaxies with similar properties (e.g., Hubble type, magnitude, optical radii and velocity); the interested reader can find details on this procedure in Persic et al. (1996) and Lapi et al. (2018). 

We mass-model the stacked RCs $v^2_{\rm model}(r)=v^2_{\rm d}(r)+v_{\rm DM}^2(r)$ as the sum of a disk component $v_{\rm d}^2(r)$
and of a DM component $v_{\rm DM}^2(r)$. We compute the gravitational potential originated by the baryons in the disk assuming standard Newtonian gravity, since we do not expect and/or require non-local effects for them; thus $v_{\rm d}^2(r)=G\,M_{\rm d}(<r)/r$ applies in terms of the disk cumulative mass $M_{\rm d}(<r)$. As customary, we assume the disk radial distribution to follow a razor-thin exponential disk (see Freeman 1970) with surface density $\Sigma_{\rm d}(r)=(M_{\rm d}/2\pi\,r_{\rm d}^2)\, e^{-r/r_{\rm d}}$ in terms of the disk scale-length $r_{\mathrm{d}}\approx r_{\mathrm{opt}}/3.2$. The related contribution to the total RC is the well-known expression $v_{\mathrm{d}}^2(r)=(G\,M_{\rm d}/r_{\rm d})\times 2\,y^{2}\,\left[I_{0}(y) K_{0}(y)-I_{1}(y) K_{1}(y)\right]$,
in terms of $y\equiv r/(2\,r_{\rm d})$ and of the modified Bessel functions $I_{0,1}$ and $K_{0,1}$. Since the fit is performed in a radial range $r\lesssim r_{\rm opt}$ we have checked that any contribution from a gaseous disk (typically more important at larger radii) is negligible and largely unconstrained, so we include only the stellar disk in the mass-modeling (se also Lapi et al. 2018; Gandolfi et al 2022b). 

The DM component of the RC is modeled in the fractional gravity framework according to $v_{\rm DM}^2(r)=|r\, {\rm d}\Phi/{\rm d}r|$ where $\Phi(r)$ is the potential given by Eqs.~(\ref{eq|pots}) and (\ref{eq|potslim}). Given the halo mass $M_{\rm H}$, we determine the halo virial radius as $R_{\rm H}\approx 260\, (M_{\rm H}/10^{12}\, M_\odot)^{1/3}$ kpc, set the concentration $c\approx 8\,(M_{\rm H}/10^{12}\,h^{-1})^{-0.1}$ according to the prescription by Dutton \& Maccio (2014), and hence determine the reference density $\rho_s=(3\, M_{\rm H}/4\pi\,R_{\rm H}^2)\times c^3/[\ln(1+c)-c/(1+c)]$ and the scale radius 
$r_s=R_{\rm H}/c$; typical values of the latter are in the range $r_s\approx 5-50$ kpc.

All in all, our model RCs depends on four parameters: the fractional index $s$, the fractional length-scale $\ell$, the halo mass $M_{\rm H}$ and the disk mass $M_{\rm d}$; in fact, we find it convenient to perform inference on the non-dimensional ratio $\ell/r_s$ instead of $\ell$. To estimate the aforementioned parameters we adopt a Bayesian MCMC technique, numerically implemented via the Python package \texttt{emcee} (see Foreman-Mackey et al. 2013). We use a standard log-likelihood $\mathcal{L}(\theta)\equiv-\chi^2(\theta)/2$ where $\theta=\{s,\log \ell/r_s,\log M_{\rm H},\log M_{\rm d}\}$ is the vector of parameters, and $\chi^2= \sum_j [v_{\rm model}(r_j|\theta)-v_{\rm obs}(r_j)]^2/\sigma_{v_{\rm obs}}^2(r_j)$ is obtained by comparing our empirical model expectations $v_{\rm model}(r_j|\theta)$ to the data $v_{\rm obs}(r_j)$ with their uncertainties $\sigma_{v_{\rm obs}}^2(r_j)$, summing over the different radii $r_j$ of each dataset (bin). 

We adopt flat priors $\pi(\theta)$ on the parameters within the ranges $s\in [1,3/2]$, $\log \ell/r_s\in [-3,1]$, $\log M_{\rm H}\, [M_\odot]\in [7,14]$, and $\log M_{\rm d}\, [M_\odot]\in [6,13]$. We then sample the posterior distribution $\mathcal{P}(\theta)\propto \mathcal{L}(\theta)\, \pi(\theta)$ by running $\texttt{emcee}$ with $10^4$ steps and $200$ walkers; each walker is initialized with a random position uniformly sampled from the (flat) priors. To speed up convergence, we adopt a mixture of differential evolution (see Nelson et al. 2014) and snooker (see ter Braak \& Vrugt 2008) moves of the walkers, in proportion of $0.8$ and $0.2$, respectively. After checking the auto-correlation time, we remove the first $20\%$ of the flattened chain to ensure the burn-in; the typical acceptance fractions of the various runs are in the range $30-40\%$. 

\section{Results} \label{sec|results}

\begin{figure}[!t]
\centering
\includegraphics[width=0.8\textwidth]{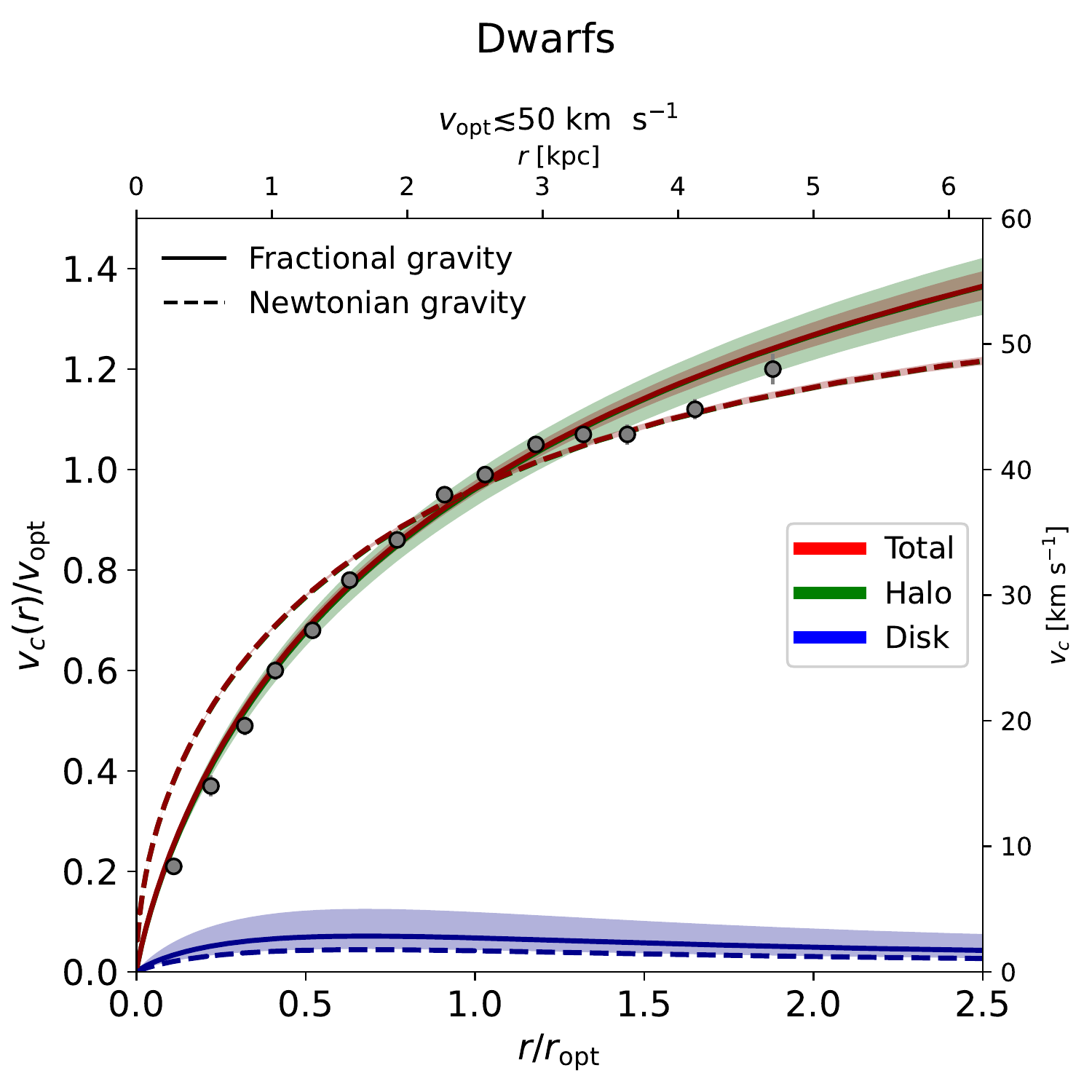}
\caption{Fit to stacked rotation curve for the dwarf galaxy sample (filled circles) with Newtonian (dashed lines) and fractional (solid lines) gravity. Red lines refer to the total rotation curve, blue line to the disk component, and green lines to the halo component. For the fractional case, the shaded areas illustrate the $1\sigma$ credible intervals from sampling the posterior distribution.}\label{fig|dwarfs}
\end{figure}

\begin{figure}[!t]
\centering
\includegraphics[width=\textwidth]{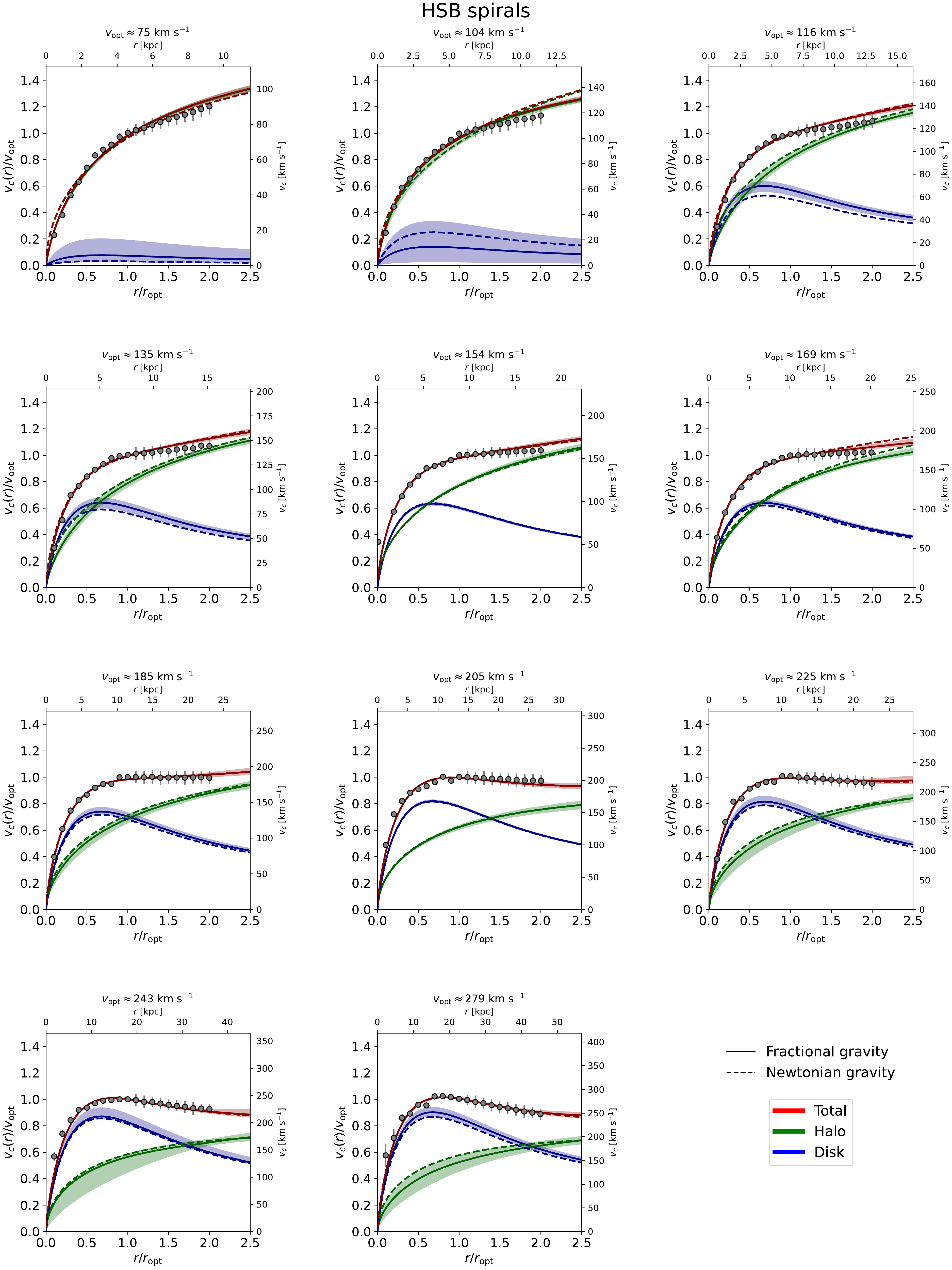}
\caption{Same as Fig. \ref{fig|dwarfs} for the stacked rotation curves of the HSB spirals sample.}\label{fig|HSB}
\end{figure}

\begin{figure}[!t]
\centering
\includegraphics[width=\textwidth]{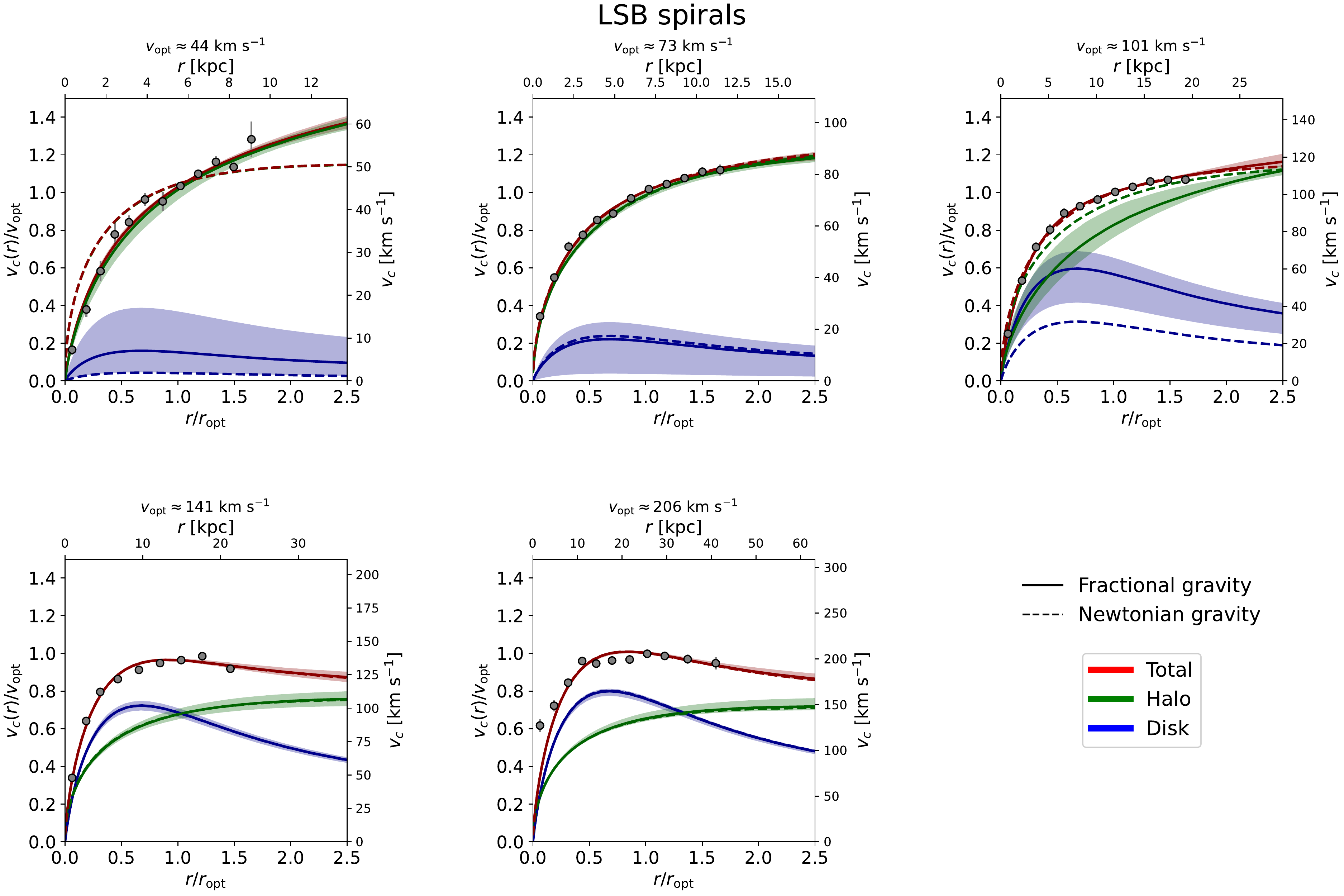}
\caption{Same as Fig. \ref{fig|dwarfs} for the stacked rotation curves of the LSB spirals sample.}\label{fig|LSB}
\end{figure}

\begin{figure}[!t]
\centering
\includegraphics[width=0.95\textwidth]{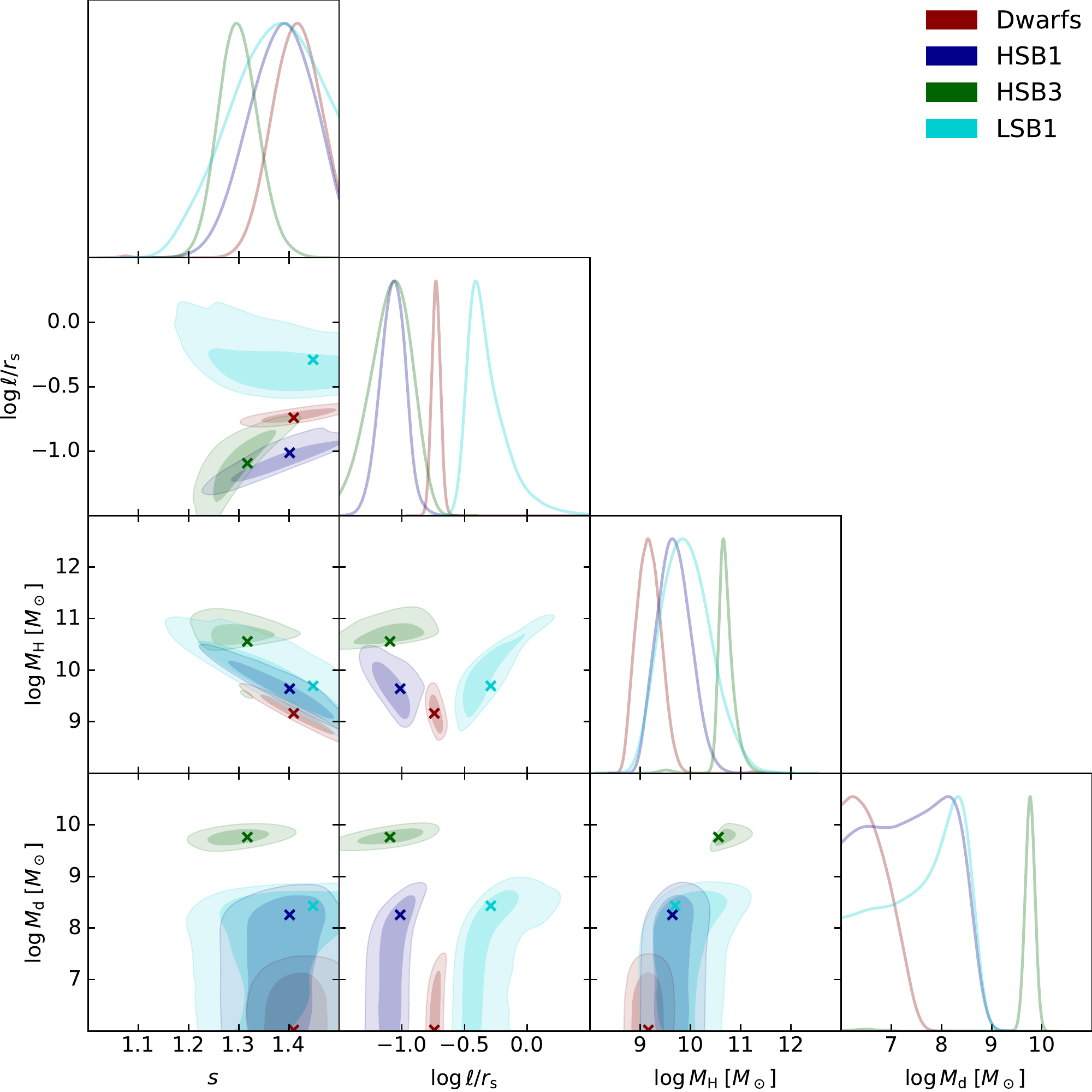}
\caption{MCMC posterior distributions in the fractional gravity framework for the fractional parameter $s$, the fractional scale-length (in units of the NFW scale radius) $\ell/r_s$, the halo mass $M_{\rm H}$ and the disk mass $M_{\rm d}$. Colored contours/lines refer to the most representative cases where there is strong evidence (from $\chi^2$ and BIC) that fractional gravity performs substantially better than the Newtonian one: red is for dwarfs, blue for HSB1 spirals, green for HSB3 spirals and cyan for LSB1 spirals. The contours show $1\sigma$ and $2\sigma$ confidence intervals, crosses mark the maximum likelihood estimates, and the marginalized distributions are in arbitrary units (normalized to 1 at their maximum value).}\label{fig|MCMC}
\end{figure}

The outcomes of the fitting procedure are illustrated in Fig. \ref{fig|dwarfs} for the dwarf sample, in Fig. \ref{fig|HSB} for HSB spirals and in Fig. \ref{fig|LSB} for LSB spirals. In each panel, the bestfit (solid lines) and the $1\sigma$ credible intervals sampled from the posterior (shaded areas) are shown for the halo (green), disk (blue), and total (red) RC models; for reference, the results in Newtonian gravity is also displayed (dashed lines). MCMC posterior distributions in the fractional gravity framework for some representative bins are shown in the cornerplot of Fig. \ref{fig|MCMC}.
Finally, Table 1 reports the marginalized posterior estimates of the fitting parameters, the reduced $\chi_r^2$ of the fits, and the difference in the Bayesian inference criterion (BIC) for model comparison between the Newtonian and fractional gravity fits\footnote{The BIC is defined as BIC $= 2\,\ln{\mathcal{L}_{\rm max}}+ N_{\rm par}\,\ln{N_{\rm data}}$ in terms of the maximum likelihood estimate $\mathcal{L}_{\rm max}$, of the number of parameters $N_{\rm par}$, and the number of data points $N_{\rm data}$. The BIC comes from approximating the Bayes factor, which gives the posterior odds of one model against another, presuming that the models are equally favored a priori. Note that what matters is the relative value of the BIC among different models; in particular, a difference of around ten or more indicates evidence in favor of the model with the smaller value.}.

The fractional gravity framework performs substantially better than the Newtonian case in several bins, especially those relative to small and intermediate mass systems, where the DM component dominates or appreciably contributes to the total RC out to $r_{\rm opt}$.  This is manifest both from the quality of the fit in terms of a substantial improvement in the reduced $\chi^2_r$, and in terms of the strongly negative values of the $\Delta$BIC (meaning that fractional gravity is favored, in a Bayesian sense, over the Newtonian case). For example, one can consider the dwarfs RC  shown in Fig. \ref{fig|dwarfs}, where the contribution of baryons to the dynamics is known to be negligible, hence the overall RC is determined solely by the DM component. In the Newtonian case, the cuspy behavior of the NFW density distribution (with standard values of the concentration) struggles in reproducing the inner part of the RC; historically, this has been one of the first evidence of the cusp-core problem. However, the dynamics in fractional gravity is substantially modified, in such a way that the RC sourced by the NFW density profile can actually reproduce the observed inner RC to a remarkable accuracy; this is because, as discussed in Section \ref{sec|methods}, the gravitational potential sourced by the NFW density in fractional gravity mirrors that sourced by a cored density distribution (e.g., a pseudo-isothermal sphere) in Newtonian gravity. In more massive galaxies, where the contribution of baryons to the inner RC out to $r_{\rm opt}$ becomes increasingly relevant and eventually dominates over the DM, the evidence for fractional gravity in the DM gets weaker; in such systems the fits in fractional and Newtonian gravity turn out to be very similar (though with an appreciable difference in the estimated total mass, see below). However, we caveat that the contribution of the DM to the RC in massive spirals emerges at radii $r\gtrsim r_{\rm opt}$ where the data are still subject to appreciable uncertainties; more precise observations over an extended radial range would be necessary before drawing a definitive conclusion on the impact of fractional gravity in massive spirals. 

\begin{figure}[!t]
\centering
\includegraphics[width=0.7\textwidth]{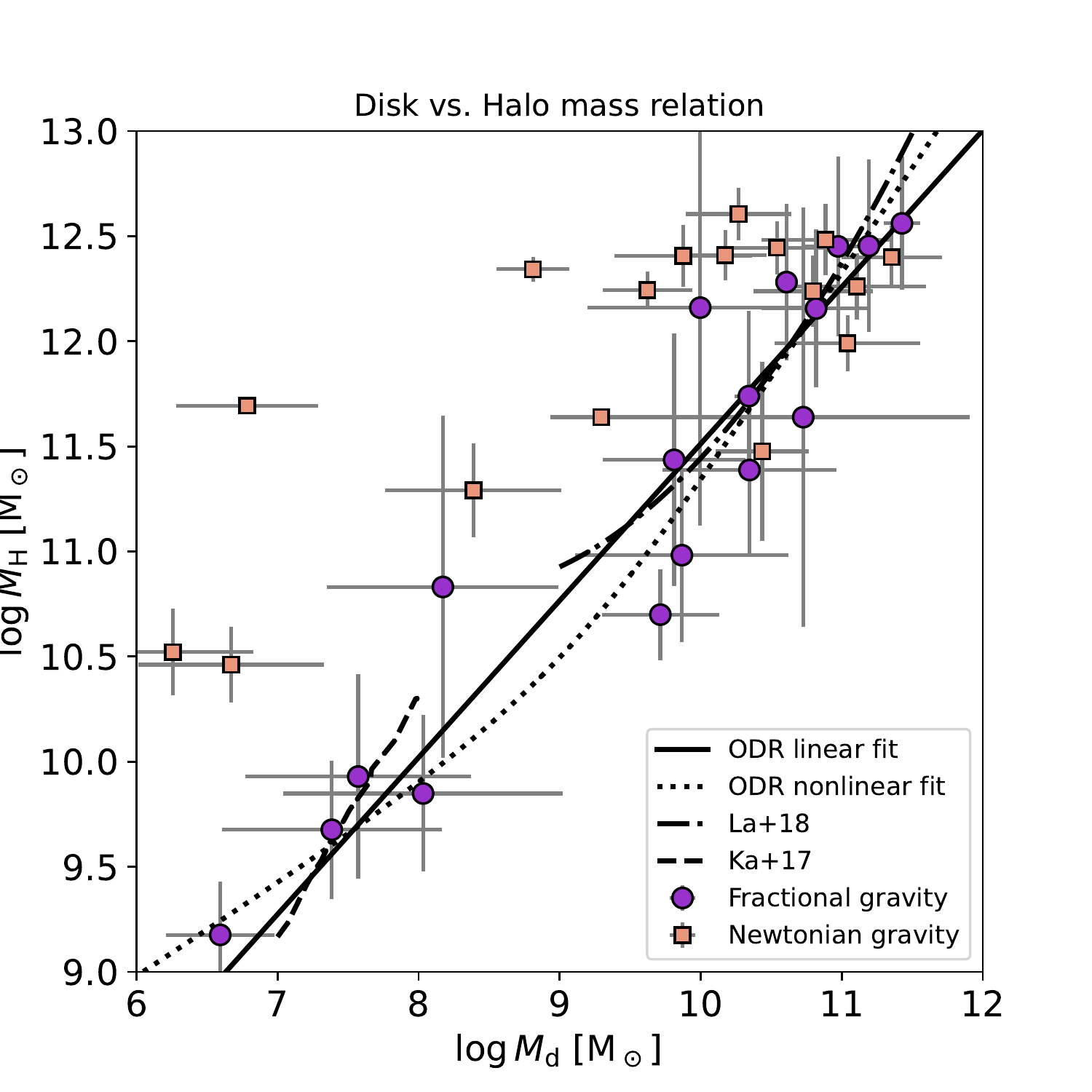}
\caption{The disk (stellar) mass vs. halo mass relation from rotation curve modeling in Newtonian (orange squares) and in fractional gravity (magenta circles). The dotted and solid lines show linear and nonlinear orthogonal regression distance fits to the fractional gravity data (see text for details). For reference, the dot-dashed line is the average relation found by Lapi et al. (2018) and the dashed line is the one by Karukes \& Salucci (2017) from the analysis of the galaxy RC with the empirical Burkert profile for the halo component.}\label{fig|SMHMR}
\end{figure}

In Fig. \ref{fig|SMHMR} we illustrate the disk (stellar) vs. halo mass relation constructed with the bestfit estimates in fractional (magenta circles) and Newtonian (orange squares) gravity. For reference, the dot-dashed and dashed lines are the relations inferred by Lapi et al. (2018) and Karukes \& Salucci (2017) by fitting galaxy RCs of different masses with the cored Burkert profile, that are also in agreement with abundance matching determinations (Moster et al. 2013; Lapi et al. 2018). The inferred masses in the fractional gravity framework are in excellent agreement with the Lapi et al. (2018) and Karukes \& Salucci (2017) results. Contrariwise, there is instead a clear tendency for the NFW fits in Newtonian gravity to yield higher DM masses at fixed stellar mass (especially so for small galaxies), that are somewhat inconsistent with the aforementioned empirical determinations; in the literature it has been pointed out that more reasonable halo masses can be obtained by letting the concentration parameter free to vary in the fit, but at the price of obtaining concentration values in disagreement with the concentration vs. mass relation measured in cosmological simulations (e.g., Dehgani et al. 2020; Gandolfi et al. 2022b). We fit the stellar mass vs. halo mass relation in fractional gravity via a orthogonal distance regression algorithm (ODR) that takes into account errorbars in both axes. Using a linear (solid line) shape  $\log M_{\rm d} [M_\odot]=a+b\,(\log M_{\rm H} [M_\odot] -11)$ we obtain bestfit parameters $a=9.31\pm 0.11$ and $b=1.34\pm 0.09$ and a reduced $\chi^2_r\approx 0.37$; a nonlinear fit (dotted line) inspired by Eq.~(2) in Moster et al. (2013) yields only a modest improvement in the reduced $\chi^2_r\approx 0.31$ and large uncertainties on the fitted parameters.

\begin{figure}[!t]
\centering
\includegraphics[width=0.49\textwidth]{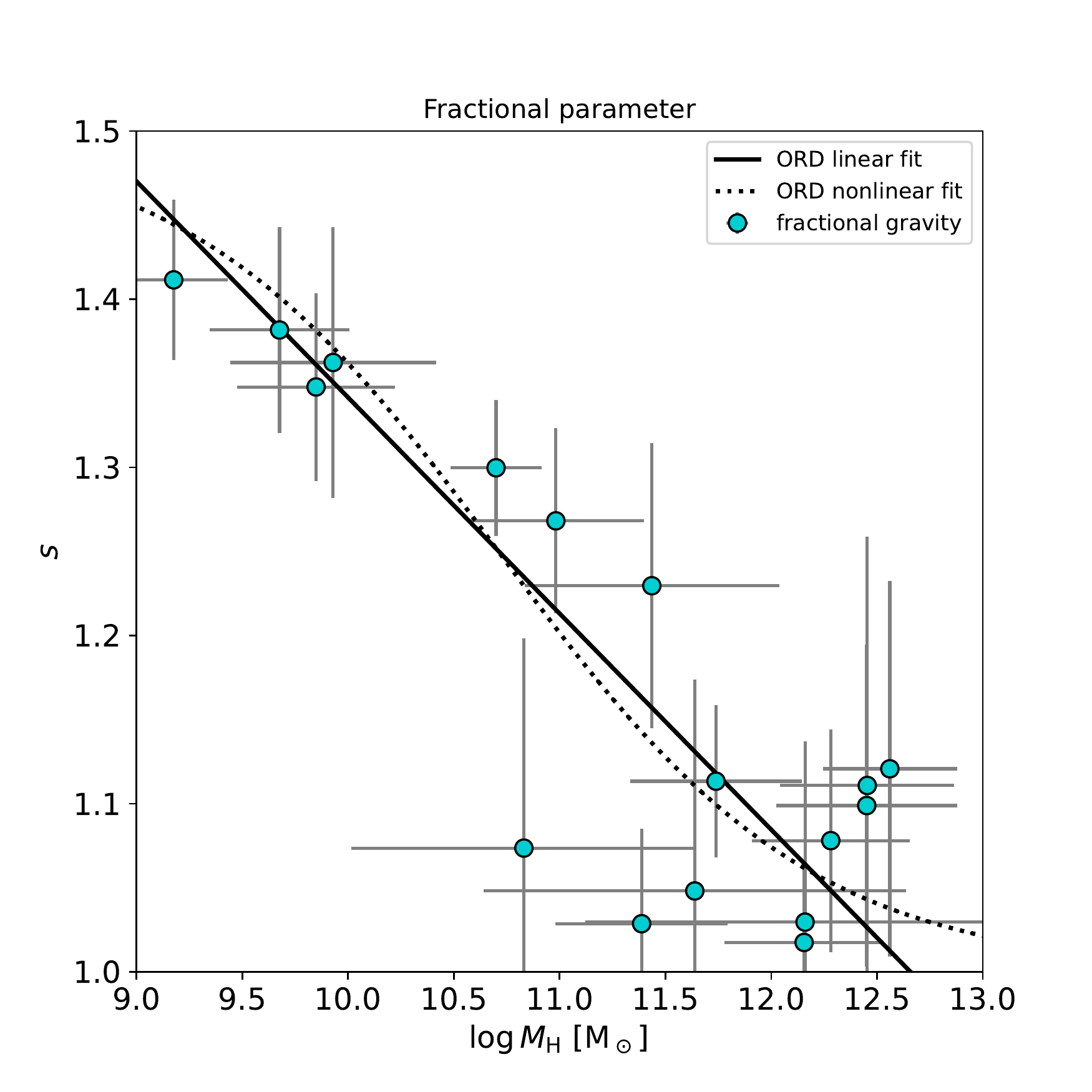}
\includegraphics[width=0.49\textwidth]{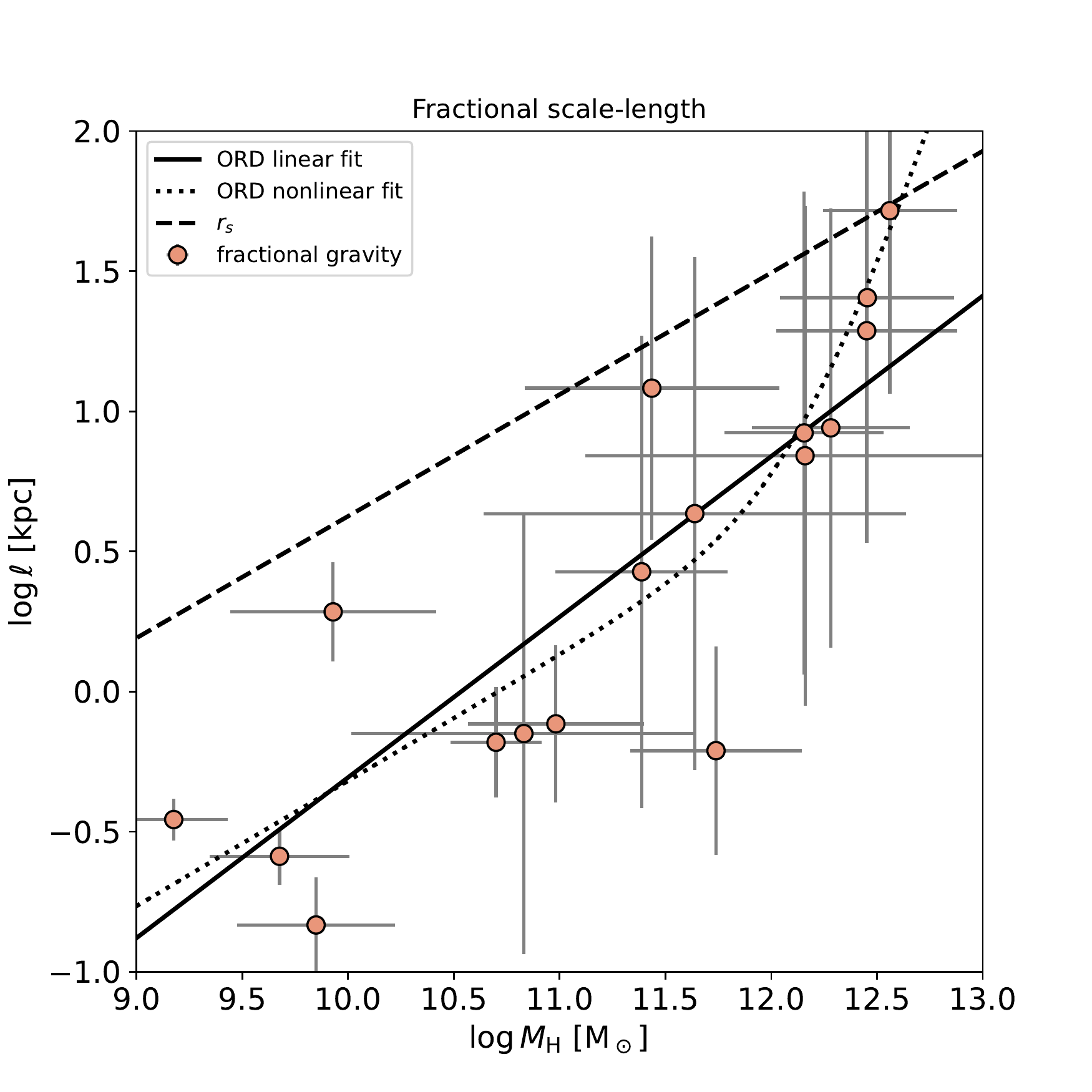}
\caption{Dependence of the parameter $s$ and of the scale-length $\ell$ in on the halo mass $M_{\rm H}$, as derived from rotation curve modeling in fractional gravity. The solid and dotted lines show linear and nonlinear orthogonal regression distance fits to the data (see text for details). In the right panel, the dashed line displays the values of the scale radius $r_s$.}\label{fig|scaling}
\end{figure}

In Fig. \ref{fig|scaling} we illustrate the behavior of the fractional gravity index $s$ and length-scale $\ell$ on halo mass. The former features a decreasing behavior as a function of halo mass, passing from values around $s\approx 1.3-1.4$ in dwarf galaxies with $M_{\rm H}\lesssim 10^{10}\, M_\odot$ to $s\approx 1.2-1.3$ in intermediate mass galaxies, to $s\lesssim 1.1$ in massive galaxies with $M_{\rm H}\gtrsim$ some $10^{11}\, M_\odot$.
This implies a strongly fractional behavior of the DM component in small and intermediate mass systems, whereas deviations from Newtonian gravity in high mass systems seems to be more limited (at least given the present quality of the data in the outer regions where the DM component can be probed in massive galaxies, see above). We fit the $s-M_{\rm H}$ relation with a linear shape  $s=a+b\,(\log M_{\rm H} [M_\odot] -11)$ via an ODR algorithm, to obtain bestfit parameters $a=1.21\pm 0.02$ and $b=-0.13\pm 0.01$ and a reduced $\chi^2_r\approx 0.58$; a nonlinear fit (dotted line) $s=\frac{5}{4}+\frac{1}{4}\,\tanh(c\,(\log M_{\rm H} [M_\odot]-d)]$
with asymptotic values $s=1$ and $1.5$ at small and large masses 
yields bestfit parameters $c=-0.68\pm 0.09$, $d=10.71\pm 0.12$
and a modest improvement in the reduced $\chi^2_r\approx 0.52$.

As to the fractional length-scale, there is a tendency of increasing with halo mass. However, it may be stressed that the determination of $\ell$ is robust only in small and intermediate mass galaxies with strong fractional gravity behavior, or in other words where the index $s$ is substantially larger than unity. Contrariwise, in massive galaxies, where $s$ gets close to $1$, the fractional Poisson equation Eq.~(\ref{eq|fracPoi}) becomes almost independent on $\ell$ (the quantity $\ell^{2s-2}$ appears on the r.h.s.), so that any robust inference on this parameter is difficult.  We fit the $\ell-M_{\rm H}$ relation with a linear shape  $\ell [{\rm kpc}]=a+b\,(\log M_{\rm H} [M_\odot] -11)$ via an ODR algorithm, to obtain bestfit parameters $a=0.27\pm 0.13$ and $b=0.57\pm 0.11$ and a reduced $\chi^2_r\approx 1.25$; a nonlinear fit (dotted line) in terms of a double powerlaw shape
leaves unchanged the reduced $\chi^2_r\approx 1.32$ and yields large errors on the fitted parameters. Interestingly, the bestfit linear relation implies a scaling of $\ell$ very similar to the NFW scale radius $r_s$, in such a way that $\ell/r_s\approx 0.25\pm 0.15$ is approximately independent of halo mass.

The trends of the parameters $s$ and $\ell$ with halo mass may be suggestive that the effects of fractional gravity on progressively larger scales become weaker; this would retain the successes of the standard DM paradigm in Newtonian gravity on cosmological scales (see Sect. \ref{sec|intro}). However, given the still large uncertainties affecting the kinematic analysis of massive galaxies, it would be interesting to validate our results by looking at larger systems like clusters of galaxies. In fact, the overall gravitational potential of the cluster is dominated by the DM component, while most of the baryons are in the form of a hot intracluster medium in hydrostatic equilibrium within the DM potential well. In fractional gravity, one may expect the thermodynamic properties of the intracluster medium (e.g., density, temperature, pressure, entropy runs) to be substantially modified, so that comparing with those empirically inferred from X-ray, Sunyaev-Zel'dovich effect and weak lensing data could be extremely informative on fractional gravity parameters at scales much larger than galaxies'. We plan to follow this program in a forthcoming paper.

\subsection{Radial Acceleration Relation}\label{sec|RAR}

The radial acceleration relation (RAR), namely a relationship between the total $g_{\rm tot}$ vs. the baryonic acceleration $g_{\rm bar}$ has been proposed by McGaugh et al. (2016) via the analysis of individual RCs of the SPARC sample (Lelli et al. 2016); note that the acceleration of any component can be simply computed from the RC as $g\equiv v^2(r)/r$. The RAR is thought to subsume and generalize a plethora of well-known dynamical laws of galaxies (see discussion by Lelli et al. 2017), although its tightness and physical interpretation are still debated (e.g., Di Paolo et al. 2019; Salucci et al. 2020). Here we just aim to show the RAR that emerges from the dynamics of galaxies in the fractional gravity framework, using as inputs the outcomes from the previous analysis of stacked RC data. 

Toward this purpose, note that the RAR is a local scaling law that combines data at different radii in galaxies with different masses, which feature different contribution of stellar/gas disk and DM. Therefore we approach the problem via a semi-empirical method: we first build up mock RCs of galaxies with different halo masses, by exploiting well known empirical relationships between halo mass and baryonic properties; then we sample the RCs to derive the total and baryonic accelerations and construct the RAR. Such a procedure has been described in Gandolfi et al. (2022b; see their Section 4), to which we defer the reader for details. The only difference with respect to that work is that the RC of the DM component is modeled in fractional gravity (with a NFW density profile). 

\begin{figure}[!t]
\centering
\includegraphics[width=0.9\textwidth]{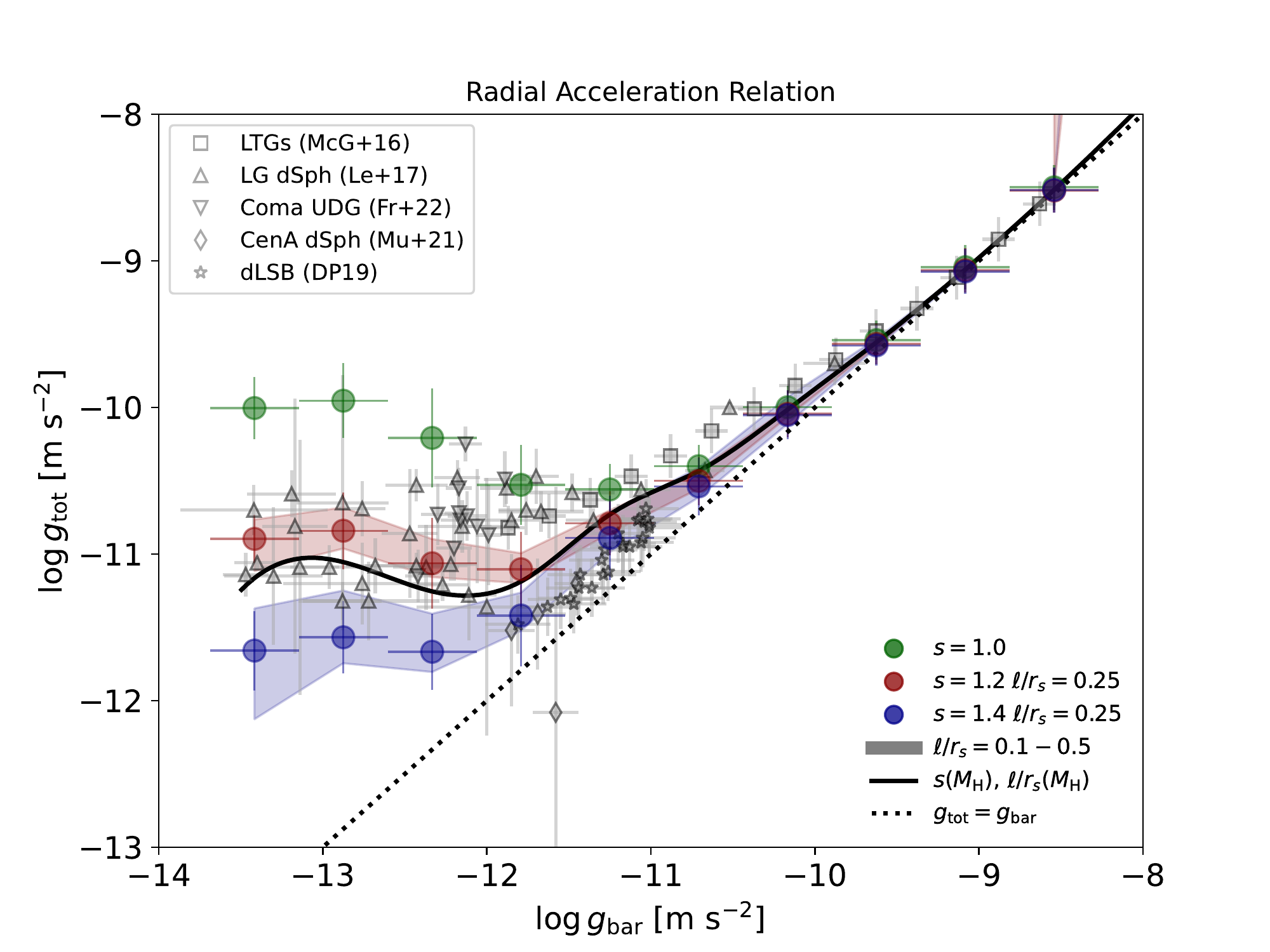}
\caption{The radial acceleration relation (RAR). Green circles illustrate the outcome for $s=1$ (independent of $\ell$ and corresponding to Newtonian gravity), red circles for $s=1.2$ and $\ell/r_s=0.25$, and blue circles for $s=1.4$ and $\ell/r_s=0.25$; the red and blue shaded area show the effect of varying $\ell/r_s$ in the range $0.1-0.5$. Finally the black curve is on adopting a halo mass-dependence in $s$ and $\ell/r_s$ as emerging from our analysis of stacked rotation curve (see Fig. \ref{fig|scaling}). For reference, the dotted black line displays the one-to-one relation $g_{\rm tot}=g_{\rm bar}$. Data for spiral galaxies (binned) are from McGaugh et al. (2016; squares), for Local Group dwarf spheroidal from Lelli et al. (2017; triangles), for Coma Cluster Ultra Diffuse Galaxies from Freundlich et al. (2022; reversed triangles), for Centaurus A dwarf spheroidals from Muller et al. (2022; diamonds), and for dwarf LSB with $g_{\rm bar}(0.4\lesssim r/R_{\rm opt}\lesssim 1)<-11$ from Di Paolo et al. (2019; stars).}\label{fig|RAR}
\end{figure}

The RAR is illustrated in Fig. \ref{fig|RAR}. The observational determination are shown in grey: data for spiral galaxies (binned) are from McGaugh et al. (2016; squares), for Local Group dwarf spheroidal from Lelli et al. (2017; triangles), for Coma Cluster Ultra Diffuse Galaxies from Freundlich et al. (2022; reversed triangles), for Centaurus A dwarf spheroidals from Muller et al. (2021; diamonds), and for dwarf LSB with $g_{\rm bar}(0.4 \lesssim r/R_{\rm opt} \lesssim 1) < -11$ from Di Paolo et al. (2019; stars). For reference, the dotted black line displays the one-to-one relation $g_{\rm tot}=g_{\rm bar}$.

The big green circles illustrate the outcome for $s=1$ (independent of $\ell$ and corresponding to Newtonian gravity); as pointed out by Gandolfi et al. (2022b) the observed RAR is reasonably reproduced at large baryonic acceleration $g_{\rm bar}$, but it is considerably overpredicted at small $g_{\rm bar}$. The big red and blue circles show the outcomes in fractional gravity for $s=1.2$ and $s=1.4$, respectively, and $\ell/r_s = 0.25$ (typical values from the stacked RC analysis); the shaded area displays instead how the predicted RAR changes when $\ell/r_s$ is varied in the range $0.1-0.5$ (extremal values from the stacked RC analysis). Finally, the solid black line is the outcome by assuming the scaling of $s$ and $\ell/r_s$ with halo mass $M_{\rm H}$ from the stacked RC analysis (see Fig. \ref{fig|scaling}). All in all, the RAR in fractional gravity has a shape and an overall scatter pleasingly consistent with the data.

These results can be understood as follows. The high acceleration regime of the RAR is mainly dominated by the contribution at small radii in high mass galaxies, where the total gravitational acceleration is anyway dominated by baryons, to imply $g_{\rm tot} \approx g_{\rm bar}$. Contrariwise, at small $g_{\rm bar}$ the behavior of the RAR is dictated by small/intermediate radii in intermediate and low mass galaxies, where the  baryon acceleration is dominated by the stellar disk, while the total gravitational acceleration is contributed by both the disk and the halo; for the NFW density profile in Newtonian gravity the halo typically dominates over the disk and the RAR deviates upward with respect to the high acceleration regime. However, in fractional gravity such an upward deviation is mitigated because the inner RC corresponding to the NFW density cusp is somewhat steepened $v_{\rm DM}\propto r^{s-1/2}\, (\ell/r_s)^{1-s}$, hence the total acceleration toward the center is appreciably smaller than in Newtonian dynamics, especially for $s$ approaching the limiting value $1.5$ and for higher $\ell/r_s$ ratios. We conclude by noticing that this behavior is different from intrinsically cored models in Newtonian dynamics, that as demonstrated by Gandolfi et al. (2022b; see their Fig. 7) tend to enforce a low-acceleration behavior similar to (or only slightly different from) the high-acceleration regime.

\subsection{Universal Core Surface Density and Core Radius vs. Disk Scale-length}\label{sec|CSF}

It has been well established (see Salucci \& Burkert 2000; Donato et al. 2009; Gentile et al. 2009; Burkert 2015) that, when fitting galaxy RCs with a cored profile like Burkert or a pseudo-isothermal sphere, the product of the core density $\rho_0$ and core radius $r_0$, which is a kind of core surface density, is an approximately universal constant among different galaxies; typical  values of the latter have been estimated around $\rho_0\, r_0\approx 140_{-30}^{+80}\, M_\odot$ pc$^{-2}$ by Donato et al. (2009), and around $\rho_0\,r_0\approx 75_{-45}^{+85}\, M_\odot$ pc$^{-2}$ by Burkert (2015). This universality in the core surface density is a somewhat unexpected property that poses a serious challenge to any theoretical model of core formation (e.g., Deng et al. 2018; Burkert 2020). We now aim to illustrate that halos in the fractional gravity framework are broadly consistent with such a remarkable scaling law.

\begin{figure}[!t]
\centering
\includegraphics[width=0.6\textwidth]{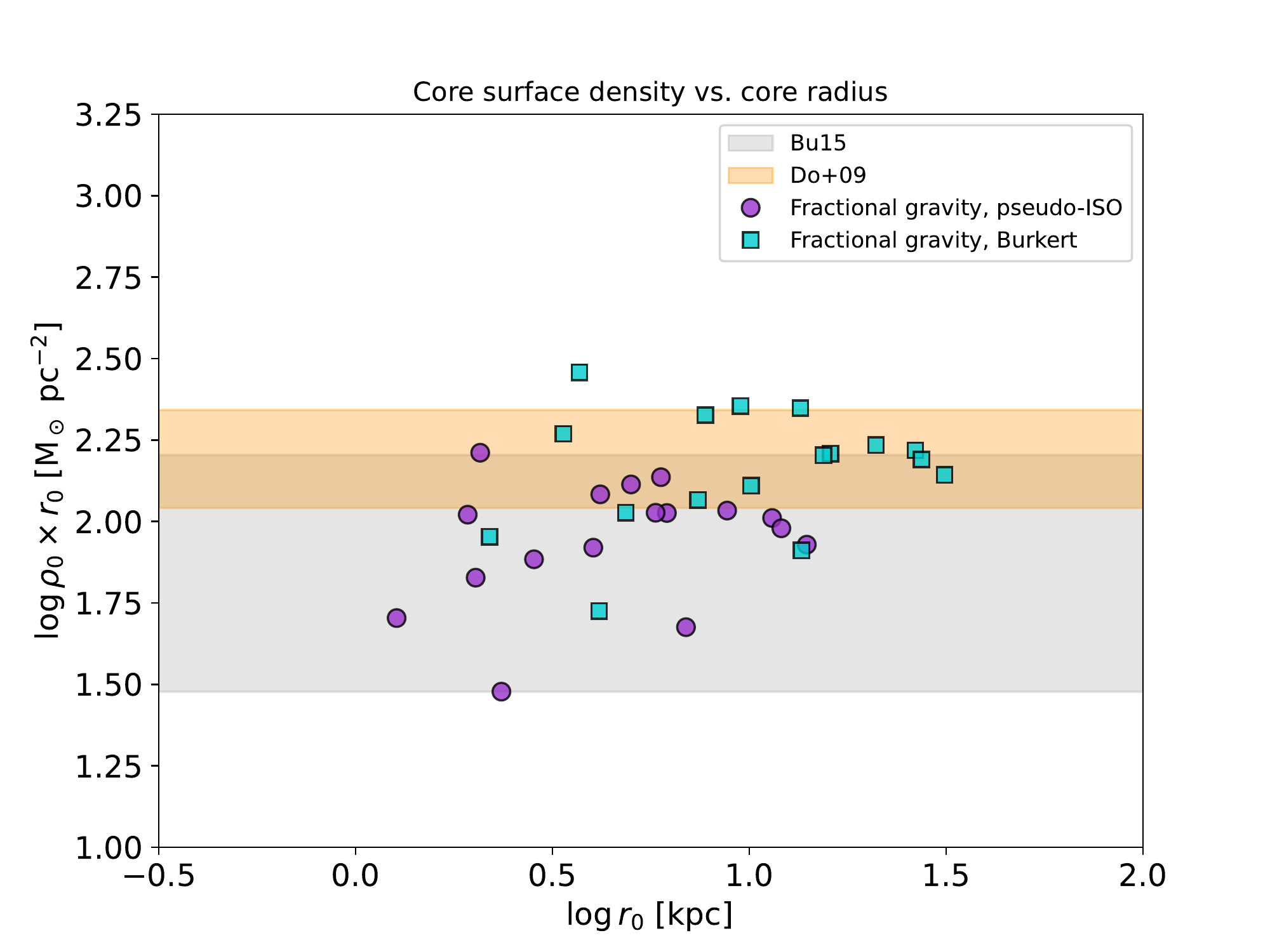}
\includegraphics[width=0.6\textwidth]{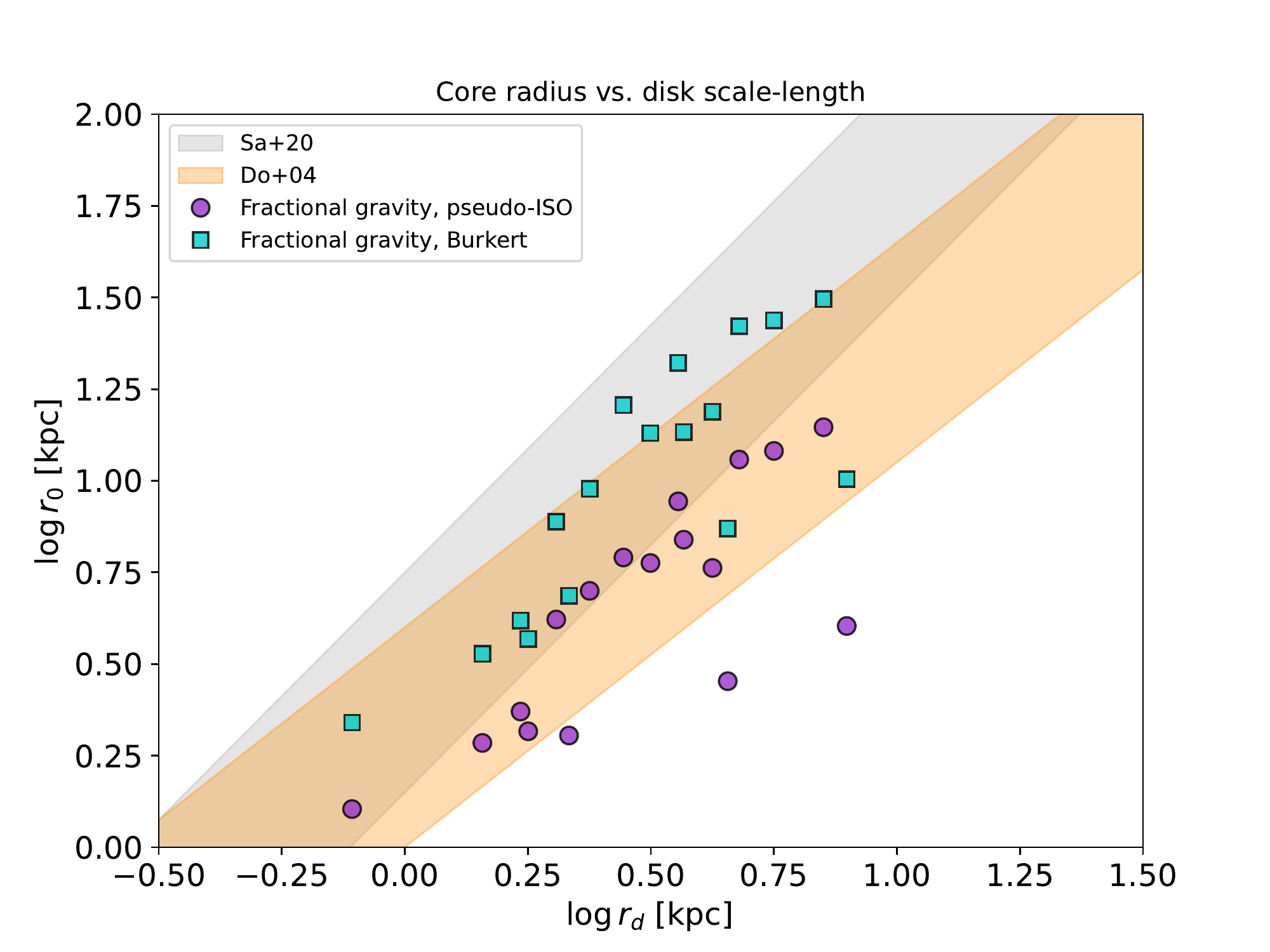}
\caption{The core surface density $\rho_0\,r_0$ vs. the core radius $r_0$ (top) and the core radius $r_0$ vs. the disk scale-length $r_{\rm d}$ (bottom). Purple circles and cyan squares illustrate the results from this work, when the core density and radius are determined by rendering our bestfit rotation curves from fractional gravity with the cored pseudo-isothermal or Burkert shapes (see text for details). In the top panel grey and orange shaded areas display the empirical scaling relations by Burkert (2015) and Donato et al. (2009), while in the bottom panel those by Salucci et al. (2020) and Donato et al. (2004).}\label{fig|CSD}
\end{figure}

To fairly compare with the literature universal core surface density relation, we need to identify an equivalent core density and core radius for a DM halo described by fractional gravity. To this purpose, we start from the bestfit rendition to the stacked RCs of Section \ref{sec|results} in the fractional gravity framework, refit each of these RCs with an equivalent Burkert or pseudo-isothermal sphere RC via a least-square minimization algorithm, and finally extract the core density and core radius. The outcome of this procedure is illustrated in Fig. \ref{fig|CSD} (top panel), where the core surface density is plotted against the core radius: magenta circles refer to the pseudo-isothermal sphere refitting, and cyan squares to the Burkert profile refitting. All in all, the measured core surface densities in the fractional gravity framework are broadly consistent with the universal relation and its overall normalization.

Another interesting relation known to hold (see Donato et al. 2004; Salucci et al. 2007; Karukes \& Salucci 2017); Lapi et al. 2018; Salucci et al. 2020) from dwarf to massive galaxies involves the core radius $r_0$ and the disk scale-length $r_{\rm d}$. Its origin is uncertain, with some of the above authors arguing for an evidence in favor of a non-gravitational interaction between baryons and DM. To check such a relationship in fractional gravity, we extract $r_0$ as described above for every stacked RC bin, and plot it as a function of the average $r_{\rm d}=r_{\rm opt}/3.2$ in the bin. The outcome is illustrated in Fig. \ref{fig|CSD} (bottom panel). We find a pleasing agreement with the determinations by Donato et al. (2004) and Salucci et al. (2020). In fractional gravity the $r_0$ vs. $r_{\rm d}$ relation can be traced back to non-local effects, that ultimately enforce the scaling with halo mass of the core radius and of the disk mass, the latter being in turn trivially related to the disk-scale-length.

\section{An alternative MONDian viewpoint?} \label{sec|MOND}

As recalled in Sect. \ref{sec|intro}, the evidence for the presence of DM in the Universe goes much beyond galaxy kinematics; thus our fiducial approach in this work has been to explore whether fractional gravity associated to the DM component could possibly alleviate some of the small-scale issues emerging from the analysis of galaxy RCs, especially dwarfs. However, for the sake of completeness and to ease the comparison with other literature studies, it is worth exploring an alternative MONDian viewpoint, where no DM component is considered, and fractional gravity is associated to the baryons. The underlying idea is that, similarly to the original version of the MOND or of other modified gravity theories, DM phenomenology could be emulated by baryons in fractional gravity.

\begin{figure}[!t]
\centering
\includegraphics[width=0.9\textwidth]{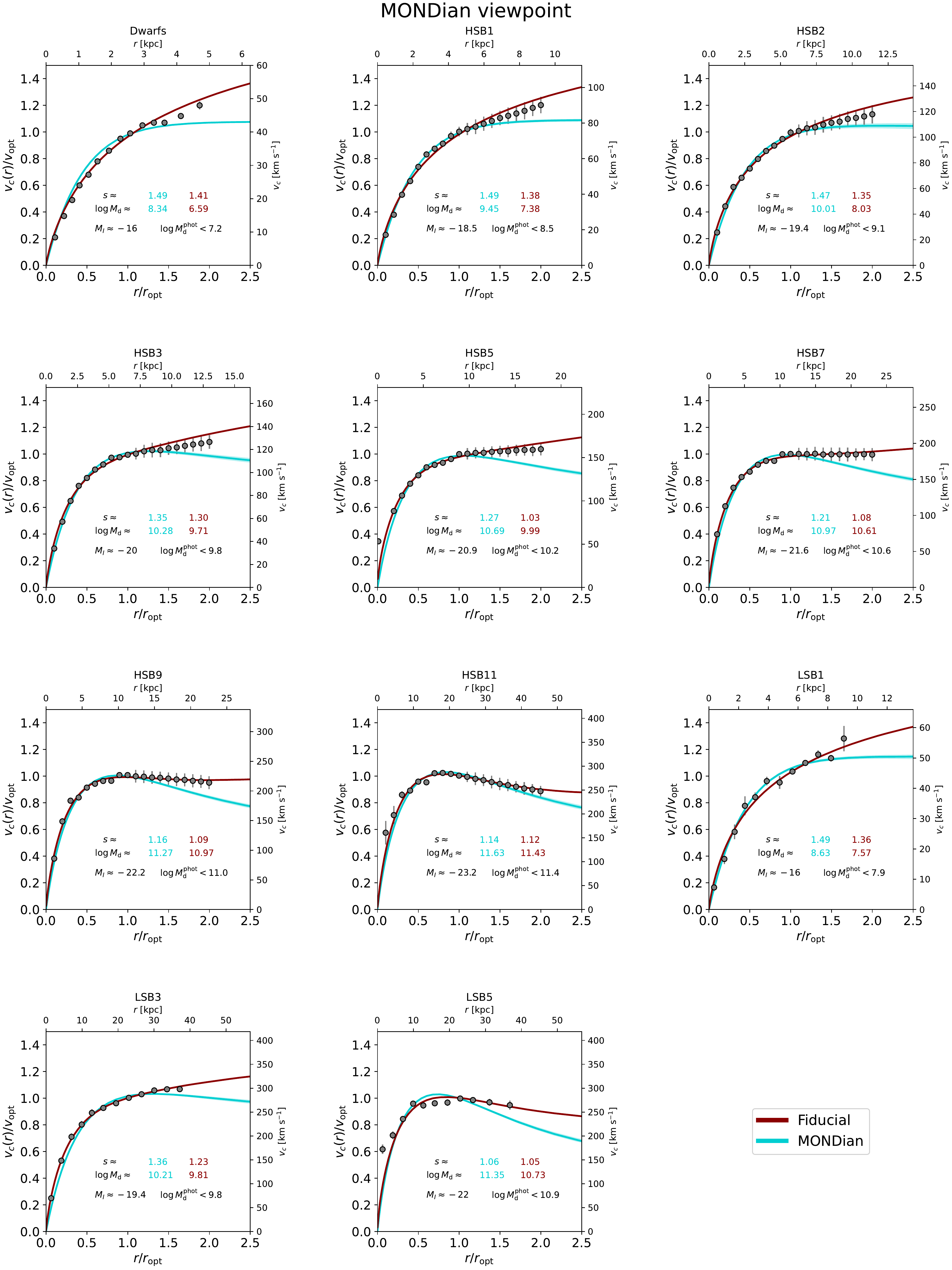}
\caption{Figure showing results for the MONDian viewpoint of fractional gravity (see Sect. \ref{sec|MOND}). For a subset of representative stacked RCs we illustrate in red the bestfit from our fiducial setup of fractional gravity with DM, and in cyan the bestfit from the MONDian setup with no DM. In each panel we also report in color the corresponding bestfit values of the parameter $s$ and $\log M_{\rm d}\,[M_\odot]$, and in black the maximal disk mass expected on the basis of the average $I-$band magnitude of the bin and typical mass-to-light-ratios, as estimated from the scaling relation by Salucci et al. (2008; see also Persic \& Salucci 1990). Note that in the MONDian viewpoint the fractional lengt-scale is related to the disk mass by $\ell\sim (2/\pi)\times \sqrt{G\, M_{\rm d}/a_0}$, and ranges from $\approx 0.3$ to $\approx 10$ kpc when moving from dwarfs to massive galaxies.}\label{fig|MOND}
\end{figure}

We have repeated the analysis of Sect. \ref{sec|Bayes} adopting this MONDian viewpoint. Specifically, we have assumed that no DM component is present, and we have fitted the stacked RCs using the circular velocity of a razor-thin exponential disk model in fractional gravity (provided in Appendix \ref{app|solutions}; we recall that for the systems considered in this work, the contribution to the stacked RCs from any bulge or gas component can be safely neglected).
In this setup the free parameters of the model are the disk mass $M_{\rm d}$, the fractional index $s$ and the fractional length-scale $\ell$. As for $\ell$, we have tried either to let it vary as a free fitting parameter, or to set it to the value $\ell\sim (2/\pi)\times \sqrt{G\, M_{\rm d}/a_0}$ with $a_0\approx 1.2\times 10^{-10}$ m s$^{-2}$ that can be naively expected in a MONDian interpretation of the fractional gravity theory (see end of Appendix \ref{app|fracalc}); we find that the fits to the stacked RCs in these two cases are practically indistinguishable, so we focus on the latter one. The results of the fitting procedure, on a subset of representative RCs of our sample, are summarized in Fig. \ref{fig|MOND}. In each panel, we illustrate  the bestfit RC in the MONDian approach with no DM, and report for comparison the one from our fiducial setup with DM; we also indicate the corresponding bestfit estimates of the fractional index $s$ and of the disk mass $M_{\rm d}$.

An important point to stress is that the MONDian viewpoint shares with our fiducial setup a similar decreasing trend of $s$, from values $s\approx 1.5$ in small galaxies toward $s\approx 1$ in the largest systems. This is essentially imposed by the behavior of the measured RCs in the inner region, which behave like a solid body in dwarfs while  are steeply rising in massive galaxies. However, it is seen that the MONDian viewpoint struggles somewhat in reproducing the outer behavior of the RCs, especially in small galaxies. This is because for $s$ appreciably larger than one, the RC of a razor-thin disk in fractional gravity (and this is true for other baryonic profiles as well, see Appendix \ref{app|solutions}) tends to saturate to a constant in the outer regions, while the measured RCs in small galaxies are quite steeply rising even in the outermost sampled point. Moreover, the fit of the MONDian viewpoint in the outer regions does not substantially improve in moving toward massive galaxies; this is because $s$ progressively decreases (as required to fit the inner part, see above), so the outer RC in fractional gravity tends to recover the steeply decreasing trend of the Newtonian case, while the measured RCs retain a rather flat or modestly decreasing behavior.

One can speculate that perhaps the MONDian viewpoint can be reconciled with the observed RCs if the fractional index $s$ is allowed to vary with the radius, to originate a sort of variable-order fractional theory (see Giusti 2000). However, some caveats are in order: first, non-trivial numerical techniques for solving the fractional Poisson equation must be developed in such a case, since the analytic treatment becomes impossible; second, it is not clear whether conceptually the variable order theory is well posed, since fractional gravity is inherently non-local, hence a simple dependence $s(r/\ell)$ on the local radial coordinate does not seem easily justified; third, to describe the radial dependence of $s$ other parameters must be introduced, which plainly make the theory subjects to additional degeneracies, and less elegant or predictive.

Another issue with the MONDian viewpoint is that, lacking the DM component, the correct normalization of the RC (even in the inner region) can be obtained only with disk masses systematically larger than in our fiducial setup. This may constitute a problem, especially in dwarf spheroidals, since the visible mass that can be inferred from photometry ($I-$band) could be appreciably smaller. In other words, the mass-to-light ratios that can be derived from the bestfit stellar mass in the MONDian viewpoint tend to be substantially larger than the typical values $M_\star/L_I\lesssim 1.5$ expected on the basis of stellar population synthesis models (e.g., Portinari et al. 2004). To highlight this point, in each panel of Fig. \ref{fig|MOND} we have reported the average $I-$band luminosity in the bin and the maximum disk mass that can be inferred from such a photometry, according to the relationship between luminosity and disk-to-total mass at the optical radius, estimated by Salucci et al. (2008; see also Persic \& Salucci 1990). There is clearly a tendency for the bestfit stellar masses in the MONDian viewpoint to exceed such values.

We also note that the values of the fractional lenght-scale derived in the MONDian viewpoint from the disk masses according to the relation $\ell\sim (2/\pi)\times \sqrt{G\, M_{\rm d}/a_0}$ range from $0.3$ kpc in dwarfs to about $10$ kpc in massive galaxies. For comparison, in our fiducial approach with DM $\ell$ ranges from $0.3$ kpc in dwarfs to a few tens kpc in massive galaxies (see right panel in Fig. \ref{fig|scaling}). Thus the overall dependence of $\ell$ on the total mass in the two approaches is rather similar within the respective fitting uncertainties.

In the context of the MONDian viewpoint, it is worth mentioning literature studies where fractional gravity is invoked as an alternative to DM in galaxies. A relevant example is the theory of Newtonian fractional-dimensional gravity by Varieschi (2020, 2021, 2022, 2023), that introduces a generalized law of Newtonian gravity in a spatial dimension $D<3$, representing the local effective Hausdorff dimension of the matter distribution. Another approach by Calcagni (2010, 2013, 2021; also Calcagni \& Varieschi 2022) relies on multi-fractional spacetimes with variable Hausdorff and spectral dimensions, directly inspired from quantum gravity theories. At variance with our fiducial setup, both these theories adopt a MONDian viewpoint where DM is not present, and the galaxy kinematics is interpreted as a pure geometrical effect. Another important difference with our approach (even in the MONDian viewpoint with no DM) is that the above theories are formulated in terms of integer-order, local operators (though generalization are possible), whereas the fractional Laplacian exploited in this study is inherently non-local. 
Recently, both Varieschi's and Calcagni's theories have been tested against a few prototypical examples of individual galaxy RCs. The former decently reproduce the measured RCs, though at the price of introducing a radially-dependent Hausdorff dimension. The latter can easily accommodate the observed rising or flattening behavior of the observed RCs in the intermediate and outer regions, while struggles somewhat in fitting the inner behavior. All in all, such theories could be promising, but an extended analysis on different mass systems, and especially dwarf galaxies with poor baryonic content would be interesting.

\section{Summary} \label{sec|summary}

In this work we have explored the possibility that the dark matter (DM) component in galaxies may originate fractional gravity; in such a framework, albeit the standard law of inertia continues to hold, the gravitational potential associated to a given DM density distribution is determined by a modified Poisson equation including fractional derivatives (i.e., derivatives of non-integer type), that accounts for non-local effects.

We have shown that in fractional gravity the dynamics of a test particle in the potential sourced by a NFW density distribution is substantially altered with respect to the Newtonian case (i.e. basing on the standard Poisson equation), mirroring what in Newtonian gravity would be originated by cored density profiles (like a pseudo-isothermal sphere). The fractional gravity framework can be fully characterized by two parameters: the index $s$ of the fractional Poisson equation, ruling the strength of the non-local effects; and a fractional length-scale $\ell$ marking the size below which gravitational effects are somewhat reduced and above which they are amplified by non-locality.

We have then tested such a model against stacked RCs of local spiral galaxies, with different properties (e.g., optical radius/velocity) and morphology (high and low surface brightness spirals, dwarfs). Our main results can be summarized as follows: 

\begin{itemize}

\item We have shown that the fractional gravity framework performs substantially better than the Newtonian case in fitting stacked RCs, especially for dwarfs and intermediate mass galaxies where the DM component dominates or appreciably contributes to the kinematics in the inner regions. 

\item Interestingly, the strength of fractional gravity effects gets progressively weaker in more massive systems dominated by baryons toward the center, though precise measurements of the outer RCs in massive spirals are needed to robustly confirm this finding. If true, this trend will imply that our fractional-gravity framework can substantially alleviate
the small-scale issues of the standard DM paradigm in Newtonian gravity, while saving
its successes on large cosmological scales.

\item We have derived the relationship between (stellar) disk and halo mass in the fractional gravity framework, finding remarkable agreement with literature determinations based on empirical RC modeling via cored profiles in Newtonian gravity.

\item We have computed the radial acceleration relation (RAR) in the fractional gravity framework, finding that its behavior at low baryonic acceleration appreciably deviates from that in Newtonian gravity, to yield an overall shape and scatter in pleasing agreement with the current observational determinations.

\item We have highlighted that the fractional gravity originates RCs that are consistent with the observed universal core surface density behavior and with the observed scaling of the core radius vs. the disk scale-length. Such relationships, which pose a serious challenge to other models of modified gravity and/or of nonstandard particle candidates, are naturally originated by the non-local nature inherent to the fractional gravity framework.

\item We have explored an alternative MONDian viewpoint on fractional gravity where the DM component is not present, and we have discussed its performances and criticalities in reproducing the observed galaxy kinematics.

\end{itemize}

All in all, the fractional gravity framework  effectively describes non-local effects in the DM component as a source of gravity, but how are these originated? Although the investigation of such an issue from the theoretical side is far beyond the scope of the present paper, in the vein of a qualitative discussion we can envisage three possibilities: (i) non-locality may stem from microscopic properties of the DM particles, e.g. some form of quantum entanglement, nonstandard interactions, or coupling with the gravitational metric; (ii) non-local dynamics may be enforced at the mesoscopic level, related to the fluid, coarse-grained description of the DM particles' collective behavior in a finite volume; (iii) non locality may be sourced on macroscopic scales by the response of a complex (e.g., clumpy and inhomogeneous) DM distribution to the long-range action of gravitational forces against DM and baryonic particles. That said, it should be noted that fractional gravity is not necessarily meant to be an ab initio theory, but may constitute an effective description for a whole class of models that imply the development of non-local effects in the DM component when sourcing the gravitational field.

In a future perspective, it will be interesting to investigate the performance of the fractional gravity framework in fitting the outermost RCs of $z\sim 1$ spiral galaxies (see Sharma et al. 2022), to more robustly check whether the strength and length-scale of the fractional dynamics show signs of evolution with redshift. In addition, our analysis of galaxy kinematics suggests that fractional gravity effects tend to fade away when moving to progressively massive systems; to confirm such a trend, a direct inspection of fractional gravity behavior on larger, cosmological scales would be welcome. A natural testbed would be provided by galaxy clusters (e.g., Gandolfi et al. 2023); specifically, we will investigate to what extent the thermodynamic profiles of the intracluster medium are affected by the modified DM gravitational potential. We also plan to revise the interpretation of strong and weak gravitational lensing events, that may be substantially altered in the fractional gravity framework. Finally, a more theoretical development will involve to apply the fractional gravity framework to non-standard DM particle candidates and check whether non-local effects may solve issues affecting these scenarios; e.g., halos in fuzzy DM have an inner core but the universal core surface density relation cannot be reproduced in Newtonian gravity (see Burkert 2020). Ultimately, it would be relevant to trace the physical origin of fractional gravity, especially in terms of the mechanism determining the index $s$ and length-scale $\ell$ in different systems and in originating the scaling of these quantities between themselves and with halo mass (so actually reducing the number of effective parameters in the theory). Finally, full $N-$body simulations could be exploited to study the emergence of the fractional gravitational dynamics in evolving cosmological structures.

\begin{acknowledgements}
We acknowledge M. Giulietti and M. Massardi for helpful discussions. We thank the anonymous referee for constructive comments. AL has been supported by the EU H2020-MSCA-ITN-2019 Project 860744 `BiD4BESt: Big Data applications for black hole Evolution STudies' and by the PRIN MIUR 2017 prot. 20173ML3WW, `Opening the ALMA window on the cosmic evolution of gas, stars and supermassive black holes'.
\end{acknowledgements}

\begin{appendix}

\section{A primer on Fractional calculus}\label{app|fracalc}

Fractional calculus is the field of mathematics dealing with differentiation and integrations of non-integer order. Nowadays it has found many applications in material science, rheology, seismology, transport phenomena, nuclear physics, medicine, finance, etc. Given that concepts and techniques related to the fractional calculus are not very common among the astrophysics community, for the reader’s convenience we present a short (non-exhaustive) primer in this Appendix. In particular, we focus on the definitions of fractional integrals and derivatives, of the fractional Laplacian, and of the fractional Poisson equation, that are often invoked and employed in the main text. 

\subsection{Fractional integrals and derivatives}

In one dimension, the simplest definition of fractional integral can be viewed as a natural extension of the well known formula (usually attributed to Cauchy)
\begin{equation}
\mathcal{I}^n f(x) = \frac{1}{(n-1)!}\,\int_0^x{\rm d}y\,f(y)\,(x-y)^{n-1}~,
\end{equation}
where $n$ is an integer and $n!=1\times 2\times...\times n$ is the factorial function. The above relates the $n-$fold  primitive of a function $f(x)$ to a convolution integral, as it can be easily verified by repeated integration by parts, provided that $f$ satisfies $f(x)=0$ for $x<0$. One is immediately led to extend the above formula for any positive real index $s$ by means of the Euler gamma function $\Gamma(s)\equiv \int_0^\infty{\rm d}x\, x^{s-1}\, e^{-x}$. In fact, since $\Gamma(n)=(n-1)!$ for integer $n$, one can define the fractional integral of order $s$ via the expression
\begin{equation}
\mathcal{I}^s f(x) = \frac{1}{\Gamma(s)}\,\int_0^x{\rm d}y\,f(y)\,(x-y)^{s-1}~,
\end{equation}
prescribing $\mathcal{I}^0 f(x)=f(x)$. The fractional integral enjoys the (semi-group) property $\mathcal{I}^s\,\mathcal{I}^p=\mathcal{I}^{s+p}$, which implies commutativity $\mathcal{I}^s\,\mathcal{I}^p=\mathcal{I}^p\,\mathcal{I}^s$ and the straightforward action on power function
\begin{equation}
\mathcal{I}^s x^a = \frac{\Gamma(a+1)}{\Gamma(a+s+1)}\,x^{a+s}~.
\end{equation}

It is natural to think that the fractional derivative could be the inverse operation of the above fractional integral; however, this idea must be implemented with some care. In particular, note that even for integer $n$ the differentiation $\mathcal{D}^n$ and integration $\mathcal{I}^n$ operators in general do not commute and $\mathcal{D}^n$ is only a left-inverse of $\mathcal{I}^n$; in fact, $\mathcal{D}^n\,\mathcal{I}^n f(x)=f(x)$ holds but the converse is not necessarily true due to the appearance of numerous integration constants. Therefore given an integer $m$ one defines the fractional derivative $\mathcal{D}^s$ of order $s\in (m-1,m)$ in such a way that $\mathcal{D}^s\equiv \mathcal{D}^m\,\mathcal{I}^{m-s}$, or more explicitly
\begin{equation}
\mathcal{D}^s f(x) = \frac{{\rm d}^m}{{\rm d}x^m}\,\left[\frac{1}{\Gamma(m-s)}\,\int_0^x{\rm d}y\, f(y)\,(x-y)^{m-s-1}\right]~,
\end{equation}
with $\mathcal{D}^0 f(x)=f(x)$. Using this definition and the semi-group property, one has 
$\mathcal{D}^s\,\mathcal{I}^s f(x) = \mathcal{D}^m\,\mathcal{I}^{m-s}\,\mathcal{I}^{s} f(x) = \mathcal{D}^m\,\mathcal{I}^{m}f(x)=f(x)$
so that $\mathcal{D}^{s}$ is truly the left inverse of $\mathcal{I}^{s}$. It is also trivial to verify the action on power functions
\begin{equation}
\mathcal{D}^s x^a = \frac{\Gamma(a+1)}{\Gamma(a-s+1)}\,x^{a-s}~.
\end{equation}

The definition of fractional derivative implies a inherently non-local behavior, since $\mathcal{D}f(x)$ does not depend only on the values of the dependent variable infinitesimally close to $x$ (as in standard calculus) but on a range of values even far away from $x$. More details on the basis of fractional calculus, including alternative definition of fractional derivatives (e.g., the Caputo one) can be found in the books by Oldham \& Spanier (1974) and Samko et al. (1993).

\subsection{Fractional Laplacian}

The standard Laplacian $\Delta$ is a linear, elliptic, second-order, local differential operator, whose applications in physics range from wave mechanics, gravity, electrodynamics, hydrodynamics, biophysics and probability theory. However, some physical phenomena have been experimentally shown to exhibit a non-local behavior, that has motivated the extension of the Laplacian to a fractional operator $(-\Delta)^s$; among the many possible definitions, here we rely on the simplest, 'spectral' one in Fourier space. In a $n-$dimensional space, the traditional Laplacian satisfies
\begin{equation}
\mathcal{F}\{-\Delta f\}(\mathbf{k}) = k^2\,\mathcal{F}\{f\}(\mathbf{k})~,
\end{equation}
where $\mathcal{F}\{\cdot\}$ denotes the Fourier transform and hereafter $k\equiv |\mathbf{k}|$. The fractional Laplacian of order $s\in (0,n/2)$ is defined as the linear operator satisfying
\begin{equation}
\mathcal{F}\{(-\Delta)^s f\}(\mathbf{k}) = k^{2s}\,\mathcal{F}\{f\}(\mathbf{k})~.
\end{equation}
Taking the inverse Fourier transform $\mathcal{F}^{-1}\{\cdot\}$ of both sides one obtains an explicit form for the fractional Laplacian in real space as
\begin{equation}
(-\Delta)^s\, f(\mathbf{x})= \mathcal{F}^{-1}\{k^{2s}\,\mathcal{F}\{f\}(\mathbf{k})\}(\mathbf{x})~.
\end{equation}

In analogy with the one-dimensional fractional integral, one may wonder whether there is an operator which acts like the inverse of the fractional Laplacian. This is provided by the Riesz operator $\mathcal{R}_s$, defined by
\begin{equation}
\mathcal{R}_s\, f(\mathbf{x})\equiv \frac{\Gamma\left(\frac{n-s}{2}\right)}{2^s\,\pi^{n/2}\,\Gamma\left(\frac{s}{2}\right)}\,\int{\rm d}^n \mathbf{x}'\,\frac{f(\mathbf{x}')}{|\mathbf{x}-\mathbf{x}'|^{n-s}}~;
\end{equation}
the above can be shown to satisfy
\begin{equation}
(-\Delta)^s\,\mathcal{R}_{2s}\,f(\mathbf{x})= \mathcal{R}_{2s}\, (-\Delta)^s\,f(\mathbf{x}) = f(\mathbf{x})
\end{equation}
so that one can naively set $\mathcal{R}_{2s}=(-\Delta)^{-s}$.

More details on the definition and properties of the fractional Laplacian can be found in the classic book by Stein (1971) and in the modern article by Lischke et al. (2020).

\subsection{Fractional Poisson equation}

In astrophysics the main relevance of the Laplacian arises in the context of Newtonian gravity; specifically, in three spatial dimensions the Poisson equation links the gravitational potential $\Phi$ to a given DM density $\rho$ as
\begin{equation}
\Delta\,\Phi(\mathbf{r}) = 4\pi G\,\rho(\mathbf{r})~.
\end{equation}
The above constitutes a partial differential equation that, when solved with appropriate boundary conditions, yields the gravitational potential anywhere in space. It is worth stressing that the Poisson equation is local in nature, meaning that the potential at a given space location depends only on the density distribution at the same point. 

The Green function $\mathcal{G}$ of the Poisson equation is the solution of the equation $\Delta\, \mathcal{G}(\mathbf{r})=4\pi G\,\delta_{\rm D}^3(\mathbf{r})$, in terms of the Dirac delta function $\delta_{\rm D}(\cdot)$; by Fourier transform this has the well-known solution $\mathcal{G}(\mathbf{r})=1/r$. The relevance of $\mathcal{G}$ is that 
the solution of the Poisson equation vanishing at infinity for a generic density distribution $\rho(\mathbf{r})$ can be written as the convolution integral
\begin{equation}
\Phi(\mathbf{r}) = \int{\rm d}^3\mathbf{r}'\,\mathcal{G}(\mathbf{r}-\mathbf{r'})\,\rho(\mathbf{r'})= - G\,\int{\rm d}^3\mathbf{r}'\,\frac{\rho(\mathbf{r'})}{|\mathbf{r}-\mathbf{r'}|}~;
\end{equation}
when applied to a point-like distribution at the origin $\rho(\mathbf{r})=m\,\delta_{\rm D}^3(\mathbf{r})$ this yields the famous potential $\Phi(\mathbf{r})=-G\, m/r$.

To introduce some non-local behavior (that may be needed to solve small-scale problem in galactic dynamics, as we advocate in the main text) one can generalize the above to the fractional case. This is accomplished by the natural ansatz
(see Giusti 2020)
\begin{equation}
(-\Delta)^s\,\Phi(\mathbf{r}) = -4\pi G\,\ell^{2-2s}\,\rho(\mathbf{r})~,
\end{equation}
where $\ell$ is a constant with physical dimensions of a length; the choice of $-\Delta$ is dictated by the need to have a positively defined operator so as to apply the properties discussed in the previous section.

The associated fractional Green function $\mathcal{G}_s$ may be found as the solution of the equation $(-\Delta)^s\,\mathcal{G}_s(\mathbf{r})=-4\pi G\,\ell^{2-2s}\,\delta_{\rm D}^3(\mathbf{r})$. Applying the Riesz operator for $s\in (0,3/2)$ one finds that 
\begin{equation}\label{eq|fracgreen}
\mathcal{G}_s(\mathbf{r}) = -4\pi G\,\ell^{2-2s}\,\mathcal{R}_{2s}\,\delta_{\rm  D}^3(\mathbf{r})= -\frac{G\,\ell^{2-2s}\,\Gamma\left(\frac{3}{2}-s\right)}{4^{s-1}\,\sqrt{\pi}\,\Gamma(s)}\,\frac{1}{r^{3-2s}}~;
\end{equation}
it is straightforward to check that $\mathcal{G}_{s=1}(\mathbf{r})$ collapses to the Newtonian Green function. The case $s=3/2$ is a bit tricky, and may be solved more conveniently in Fourier space; there the Poisson equation reads $\mathcal{F}\{\mathcal{G}_s\}(\mathbf{k})=-4\pi G/\ell\,k^3$ and thus $\mathcal{G}_s=-(4\pi G/\ell)\times \mathcal{F}^{-1}\{1/k^{3}\}$. Given that $-k^2\,\mathcal{F}\{\ln(r/\ell)\}=\mathcal{F}\{\Delta\log(r/\ell)\}=\mathcal{F}\{1/r^2\}=2\pi^2/k$ (where the last equality is meant to hold in the principal value sense) one finds $\mathcal{F}\{\ln(r/\ell)\}=-2\pi^2/k^3$ and hence
\begin{equation}\label{eq|fracgreenlim}
\mathcal{G}_{s=3/2}(\mathbf{r}) = \frac{2 G}{\pi\,\ell}\, \ln\left(\frac{r}{\ell}\right)~.
\end{equation}
The potential for a point mass $m$ located at the origin goes like $\Phi_s(\mathbf{r})\propto G\,m\,\ell^{2-2s}/r^{3-2s}$ for $s\in (0,3/2)$ and like $\Phi_{s=3/2}(\mathbf{r})\propto (G\,m/\ell)\times \ln (r/\ell)$ for $s=3/2$, so decreasing faster than the Newtonian case ($s=1$) for $s\in (0,1)$ and slower for $s\in (1,3/2]$. 

Correspondingly, the rotational velocity $v^2(r)=r\, |{\rm d}\Phi/{\rm d}r|$ of a test mass on a circular orbit in such a potential goes like $v_s(r)\propto \sqrt{G m/\ell}\times (\ell/r)^{3/2-s}$. Curiously for $s=3/2$ it is found that $v_{s=3/2}(r)=\sqrt{2\, G\, m/\pi\,\ell}$ is constant, an occurrence that has triggered some connection with MOND theories (see Giusti 2020); in fact, since at low accelerations MOND implies that $v^4=G\,m\,a_0$ in terms of a universal constant $a_0\approx 1.2\times 10^{-10}$ m s$^{-2}$, the two expressions are consistent provided that $\ell =  (2/\pi)\times \sqrt{G\,m/a_0}$ applies. We are not adopting this MONDian viewpoint as our fiducial setup in the present paper, but we explore it somewhat in Sect. \ref{sec|MOND}.

\section{Solutions of the fractional Poisson equation}\label{app|solutions}

In this Appendix we provide original analytic solutions $\Phi_s(r)$ of the fractional Poisson equation for various literature density profiles in spherical symmetry. 
To derive these, it is convenient to use coordinates $\mathbf{r}'=(r',\theta,\phi)$, placing the point $\mathbf{r}$ where the potential is calculated at the north pole; using Eq.~(\ref{eq|fracgreen}) for $s\in [1,3/2)$ one has
\begin{equation}
\Phi_s(r)=\int{\rm d}^3 \mathbf{r}'\,\mathcal{G}_s(\mathbf{r}-\mathbf{r}')\, \rho(r')=-\frac{G\,\ell^{2-2s}\,\Gamma\left(\frac{3}{2}-s\right)}{4^{s-1}\,\sqrt{\pi}\,\Gamma(s)}\int_0^{2\pi}{\rm d}\phi\,\int_0^\infty{\rm d}r'\,r'^2\,\rho(r')\,\int_0^\pi{\rm d}\theta\, \frac{\sin\theta}{(r^2+r'^2-2\,r\,r'\,\cos\theta)^{3/2-s}}~.
\end{equation}
A straightforward calculation yields the solution
\begin{equation}\label{eq|fracpotsphe}
\Phi_s(r)=-\frac{\sqrt{\pi}\,G\,\ell^{2-2s}\,\Gamma\left(\frac{3}{2}-s\right)}{4^{s-3/2}\,(2s-1)\,\Gamma(s)}\, \frac{\mathcal{J}_s^\rho(r)}{r}~,
\end{equation}
where
\begin{equation}
\mathcal{J}^\rho_s(r)=\int_0^\infty{\rm d}r'\, r'\, \rho(r')\, [(r+r')^{2s-1}-|r-r'|^{2s-1}]~.
\end{equation}

In the case $s=3/2$ one has instead to use Eq.~(\ref{eq|fracgreenlim}), obtaining
\begin{equation}
\Phi_{s=3/2}(r)=\int{\rm d}^3 \mathbf{r}'\,\mathcal{G}_{s=3/2}(\mathbf{r}-\mathbf{r}')\, \rho(r')=\frac{2 G}{\pi\ell}\,\int_0^{2\pi}{\rm d}\phi\,\int_0^\infty{\rm d}r'\,r'^2\,\rho(r')\,\int_0^\pi{\rm d}\theta\, \sin\theta\,\ln\left(\frac{\sqrt{r^2+r'^2-2\,r\,r'\,\cos\theta}}{\ell}\right)~,
\end{equation}
which yields the integral representation
\begin{equation}
\Phi_{s=3/2}(r)=\frac{2 G\, \ell}{r}\,\int_0^\infty{\rm d}r'\,r'\,\rho(r')\,\left[\left(\frac{r+r'}{\ell}\right)^2\,\ln\left(\frac{r+r'}{\ell}\right)-\left(\frac{r-r'}{\ell}\right)^2\,\ln\left(\frac{r-r'}{\ell}\right)-2\,\frac{r\,r'}{\ell^2}\right]~.
\end{equation}

Another method of solving the fractional Poisson equation for spherically-symmetric distribution is to work in the Fourier domain, where the density can be represented as
\begin{equation}
\mathcal{F}\{\rho\}(k)=\frac{4\pi}{k}\,\int_0^\infty{\rm d}r\, r\,\rho(r)\,\sin(k r)~.
\end{equation}
Then by inverse Fourier transform the fractional potential for $s\in [1,3/2]$ is given by
\begin{equation}\label{eq|fracpotFou}
\Phi_{s}(r)=-\frac{2 G\,\ell^{2-2s}}{\pi\, r}\, \int_0^\infty{\rm d}k\,\frac{\sin(k r)}{k^{2s-1}}\,\mathcal{F}\{\rho\}(k)~;
\end{equation}
in particular, such a Fourier method turns out to be very useful to solve the $s=3/2$ case. 

Below we will derive the solutions for the following density profiles: NFW, Hernquist, Plummer and razor-thin exponential disk. Note that in the main text we mainly consider the NFW and exponential disk density profiles; however, the others solutions are presented for completeness, and for the possible exploitation from the interested scientific community.

\subsection{Navarro-Frenk-White profile}

The Navarro-Frenk-White (NFW; Navarro et al. 1997) density profile, which is routinely exploited to model the DM distribution within virialized halos, is defined as
\begin{equation}
\rho(r) = \frac{\rho_s\,r_s^3}{r\,(r+r_s)^2}~,
\end{equation}
in terms of a scale density $\rho_s$ andf of a scale radius $r_s$; note that the cumulative mass is logarithmically divergent at large distances, so a finite halo boundary must be introduced (usually taken to be the virial radius). 
The Newtonian potential can be easily computed
\begin{equation}
\Phi_{s=1}(r) = -\frac{4\pi G\,\rho_s\,r_s^3}{r}\,\log\left(1+\frac{r}{r_s}\right)~.
\end{equation}

In the fractional case, one has to solve Eq.~(\ref{eq|fracpotsphe}) in terms of the integral
\begin{equation}
\mathcal{J}^\rho_s(r)=\rho_s\,r_s^3\,\int_0^\infty{\rm d}r'\, \frac{(r+r')^{2s-1}-|r-r'|^{2s-1}}{(r'+r_s)^2}~,
\end{equation}
that converges for $s\in [1,3/2)$ to
\begin{equation}
\begin{aligned}
\Phi_{s}(r) = & -\sqrt{\pi}\,G\,\rho_{s}\,r_s^{2}\,\left(\frac{2\,\ell}{r_s}\right)^{2-2s}\,\frac{\Gamma\left(\frac{3}{2}-s\right)}{\Gamma(s+1)}\,\frac{r_s}{r}\,\left\{\frac{2\pi s}{\sin(2\pi s)}\,\left[\left(1+\frac{r}{r_s}\right)^{2s-2} - \left(1-\frac{r}{r_s}\right)^{2s-2}\right]+\right.\\
& \\
& +\left. \frac{(r/r_s)^{2s}}{1-(r/r_s)^{2}}\,\left[\left(1+\frac{r}{r_s}\right)\, _{2}F_{1}\left(1,1,2s+1,\frac{r}{r_s}\right) + \left(1-\frac{r}{r_s}\right)\, _{2} F_{1}\left(1,1,2s+1,-\frac{r}{r_s}\right)-\frac{4s}{2s-1} \right] \right\}~;
\end{aligned}
\end{equation}
in the above ${}_2F_1$ is the ordinary hypergeometric function, which is defined by the power series ${}_2F_1(a,b,c;x)\equiv \sum_{k=0}^\infty (a)_k\,(b)_k\,x^k/(c)_k\,k!$ in terms of the Pochammer's symbols $(q)_0=1$ and $(q)_k=q\,(q+1)\,\ldots\,(q+k-1)$ for any positive integer $k$.

The case $s=3/2$ is more easily treated in the Fourier domain, and requires regularizing the integrals appearing in Eq.~(\ref{eq|fracpotFou}); the result reads
\begin{equation}
\begin{aligned}
\Phi_{s=3/2}(r) =& -\frac{4\,G\,\rho_{s}\,r_s^3}{\ell}\,\frac{r_s}{r}\,	
\left\{2\,\frac{r}{r_s}\, \left[\log \left(\frac{r}{r_s}\right)-1\right]-\left(1+\frac{r}{r_s}\right)\,\log \left(\frac{r}{r_s}\right)\,\log \left(1+\frac{r}{r_s}\right) + \right.\\
&\\
&  +\left. \left(\frac{r}{r_s}-1\right)\, \text{Li}_2\left(1-\frac{r}{r_s}\right)-\left(1+\frac{r}{r_s}\right)\, \text{Li}_2\left(-\frac{r}{r_s}\right) + \frac{\pi^{2}}{6}\right\}~,
\end{aligned}
\end{equation}
in terms of the dilogarithm function ${\rm Li}_2(x)\equiv \sum_{k=1}^\infty\, x^k/k^2$.

The fractional NFW potential will be extensively exploited in the main text (see Section \ref{sec|methods}); the corresponding (suitably normalized) rotation curve $v^2=r\,|{\rm d}\Phi/{\rm d}r|$ is illustrated in Fig. \ref{fig|model}.  

\subsection{Hernquist profile}
	
The Hernquist (1990) density profile, which is often used to model the stellar distribution within galactic bulges, is defined as
\begin{equation}
\rho(r) = \frac{M_\infty}{2\pi}\,\frac{a}{r\,(r+a)^{3}}~,
\end{equation}
in terms of a scale radius $a$ and of the total mass $M_\infty$. 
The Newtonian potential can be easily computed
\begin{equation}
\Phi_{s=1}(r) = -\frac{G M_\infty}{a}\,\frac{1}{1+r/a}~.
\end{equation}

In the fractional case, one has to solve Eq.~(\ref{eq|fracpotsphe}) in terms of the integral
\begin{equation}
\mathcal{J}^\rho_s(r)=\frac{M_\infty\,a}{2\pi}\,\int_0^\infty{\rm d}r'\, \frac{(r+r')^{2s-1}-|r-r'|^{2s-1}}{(r'+a)^3}~,
\end{equation}
that converges for $s\in [1,3/2)$ to
\begin{equation}
\begin{aligned}
\Phi_{s}(r) = & -\frac{G M_\infty}{a}\,\frac{\Gamma\left(\frac{3}{2}-s\right)}{4^{s-1}\,\sqrt{\pi}\,(2s-1)\, \Gamma(s)}\,\left(\frac{\ell}{a}\right)^{2-2s}\, \left\{\frac{(2s-3)\, (a/r) + (2s-1)\, (r/a)}{(1-r^{2}/a^{2})^{2}}\, \left(\frac{r}{a}\right)^{2s}+\right.\\
&\\
& + \left. \frac{\pi\, (2s-1)(s-1)}{\sin(2\pi s)}\, \frac{a}{r}\, \left[\left(1-\frac{r}{a}\right)^{2s-3} - \left(1+\frac{r}{a}\right)^{2s-3}\right]-\frac{(2s-1)\,(s-1)}{2\,s\, (1-r^{2}/a^{2})^{2}}\, \left(\frac{r}{a}\right)^{2s-1}\times \right.\\
& \\
& \left. \times \left[\left(1+\frac{r}{a}\right)^{2}\;_{2}F_{1}\left(1,1,2s+1,\frac{r}{a}\right) + \left(1-\frac{r}{a}\right)^{2}\;_{2}F_{1}\left(1,1,2s+1,-\frac{r}{a}\right) \right] \right\}~.\\
\end{aligned}
\end{equation}
The case $s=3/2$ is more easily treated in the Fourier domain, and requires regularizing the integrals appearing in Eq.~(\ref{eq|fracpotFou}); the result reads
\begin{equation}
\Phi_{s=3/2}(r) = \frac{G M_\infty}{\ell}\,\frac{2}{\pi}\,\left\{\ln\left(\frac{r}{a}\right) + \frac{\pi^{2}}{12}\,\frac{a}{r} -\frac{1}{2}\,\frac{a}{r}\,\left[\ln\left(\frac{r}{a}\right)\,\ln\left(1+\frac{r}{a}\right)+{\rm Li}_{2}\left(-\frac{r}{a}\right)+{\rm Li}_{2}\left(1-\frac{r}{a}\right)\right]\right\}~.
\end{equation}

The (suitably normalized) rotation curve $v^2=r\,|{\rm d}\Phi/{\rm d}r|$ corresponding to the Hernquist profile is illustrated in Fig. \ref{fig|auxprofiles} (left panel).  

\subsection{Plummer profile}

The Plummer (1911) density profile, which is often exploited to model the stellar distribution within globular clusters, is defined as
\begin{equation}
\rho(r) = \frac{3\,M_\infty}{4\pi\,a^3}\,\left(1+\frac{r^2}{a^2}\right)^{-5/2}~,
\end{equation}
in terms of a scale radius $a$ and of the total mass $M_\infty$. 
The Newtonian potential can be easily computed
\begin{equation}
\Phi_{s=1}(r) = -\frac{G M_\infty}{a}\,\frac{1}{\sqrt{1+r^2/a^2}}~.
\end{equation}

The fractional Poisson equation is most easily solved in Fourier space via Eq.~(\ref{eq|fracpotFou}); for $s\in [1,3/2)$ we obtain the expression
\begin{equation}
\begin{aligned}
\Phi_{s}(r) = & \frac{G M_\infty}{a}\,\frac{\Gamma(\frac{3}{2}-s)\,\Gamma(\frac{5}{2}-s)}{2^{2s-1}\,\pi\,(s-1)\,(s-2)}\,\left(\frac{\ell}{a}\right)^{2-2s}\,\left\{\left(1+\frac{a^2}{r^2} \right)\;_{2}F_{1}\left(\frac{3}{2}-s,\frac{5}{2}-s,-\frac{1}{2},-\frac{r^2}{a^2}\right)+\right.\\
& \\
& -\left. \left[4\,(2-s)+\frac{a^2}{r^2}\right]\;
_{2}F_{1}\left(\frac{3}{2}-s,\frac{5}{2}-s,+\frac{1}{2},-\frac{r^2}{a^2}\right) \right\}~.\\
\end{aligned}
\end{equation}
The case $s=3/2$, found after appropriate regularization, reads
\begin{equation}
\Phi_{s=3/2}(r) = \frac{G M_\infty}{\ell}\,\frac{2}{\pi}\, \sqrt{1+\frac{a^2}{r^2}}\,{\rm arcsinh}\left(\frac{r}{a}\right)~.
\end{equation}

The (suitably normalized) rotation curve $v^2=r\,|{\rm d}\Phi/{\rm d}r|$ corresponding to the Plummer profile is illustrated in Fig. \ref{fig|auxprofiles} (middle panel).  

\subsection{Razor-thin exponential disk}

The density profile of a razor-thin exponential disk in cylindrical coordinates is defined as
\begin{equation}
\rho(R,z) = \frac{M_\infty}{2\pi\,r_{\rm d}^2}\,e^{-R/r_{\rm d}}\,\delta_{\rm D}(z)
\end{equation}
The Newtonian potential on the plane of the disk ($z=0$) reads
\begin{equation}
\Phi_{s=1}(R,0) = -\frac{G M_\infty}{r_{\rm d}}\,\frac{R}{2\,r_{\rm d}}\,\left[I_0\left(\frac{R}{2\,r_{\rm d}}\right)\,K_1\left(\frac{R}{2\,r_{\rm d}}\right)-I_1\left(\frac{R}{2\,r_{\rm d}}\right)\,K_0\left(\frac{R}{2\,r_{\rm d}}\right)\right]~.
\end{equation}

The solution of the fractional Poisson equation may be most easily found in Fourier space after Eq.~(\ref{eq|fracpotFou}); for an index $s\in [1,3/2)$ we find
\begin{equation}
\begin{aligned}
\Phi_{s}(R,0) = &  -\frac{G M_\infty}{r_{\rm d}}\, \frac{\Gamma\left(s-1/2 \right)}{2\,\sqrt{\pi}\,\Gamma(s)}\, \left(\frac{\ell}{r_{\rm d}}\right)^{2-2s}\, \left[2^{2s}\, \frac{\Gamma(2-2s)\, \Gamma(s)}{\Gamma(1-s)}\; _{1}F_{2}\left(\frac{3}{2}-s;1,1-s,\frac{R^2}{4\,r_{\rm d}^2}\right)+ \right. \nonumber \\
& \\	
& - \left. \frac{\pi}{\sin(\pi s)\, \Gamma^2(s+1)}\left(\frac{R}{2\,r_{\rm d}}\right)^{2s}\;_{1}F_{2}\left(\frac{3}{2};1+s,1+s,\frac{R^2}{4\,r_{\rm d}^2}\right)\right]~;
\end{aligned}
\end{equation}
in the above ${}_1F_2$ is the generalized hypergeometric function, which is defined by the power series ${}_1F_2(a,b,c;x)\equiv \sum_{k=0}^\infty (a)_k\,x^k/(b)_k\,(c)_k\,k!$ in terms of the Pochammer's symbols $(q)_0=1$ and $(q)_k=q\,(q+1)\,\ldots\,(q+k-1)$ for any positive integer $k$.

In the case $s=3/2$, after regularization of the integrals, we obtain
\begin{equation}
\Phi_{s=3/2}(R,0) = -\frac{G\, M_\infty}{\ell}\,  \frac{2}{\pi}\,\left[{\rm Ei}\left(-\frac{R}{r_{\rm d}}\right)-e^{-R/r_{\rm d}}-\ln\left(\frac{R}{r_{\rm d}}\right)\right]~,
\end{equation}
in terms of the exponential integral function ${\rm Ei}(x)\equiv \int_{-\infty}^x{\rm d}t\, e^t/t$.

\begin{figure}[!t]
\figurenum{B1}
\centering
\includegraphics[width=\textwidth]{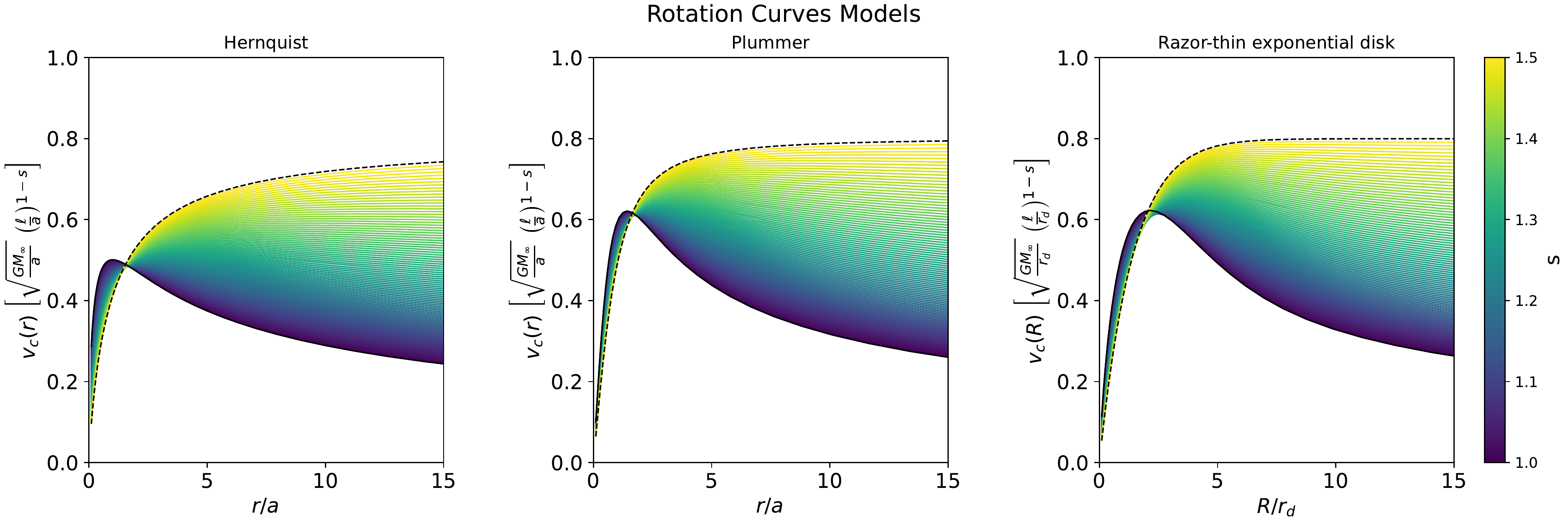}
\caption{Rotation curve models corresponding to the Hernquist (left), Plummer (middle) and razor-thin exponential disk (right) profiles for different values of the fractional parameter $s$ (color-coded). For reference, the dashed line refers to the maximal value $s=3/2$ yielding an asymptotically flat velocity.}\label{fig|auxprofiles}
\end{figure}

The (suitably normalized) rotation curve $v^2=R\,|{\rm d}\Phi/{\rm d}R|$ corresponding to the razor-thin exponential disk profile is illustrated in Fig. \ref{fig|auxprofiles} (right panel).  

\end{appendix}

\begin{deluxetable*}{cccccccccccccccccccccc}
\tablecaption{Marginalized posterior estimates (mean and $1\sigma$ confidence intervals) for the fractional and Newtonian gravity. Columns report the values of: fractional parameter $s$, fractional scale-length $\ell$, halo mass $M_{\rm H}$, disk mass $M_{\rm d}$, reduced $\chi_r^2$. The last column refer to the difference in values of the Bayesian inference criterion (BIC) between the two models.}
\tablewidth{0pt}
\tablehead{& & \multicolumn{5}{c}{Fractional gravity} & & \multicolumn{3}{c}{Newtonian gravity} & &\\
\cline{3-7} \cline{9-11} \\
\colhead{Bin} & & \colhead{$s$} & \colhead{$\log \ell/r_s$} & \colhead{$\log M_{\rm H}$ [M$_\odot$]} & \colhead{$\log M_{\rm d}$ [M$_\odot$]} & \colhead{$\chi^2_r$} & & \colhead{$\log M_{\rm H}$ [M$_\odot$]} & \colhead{$\log M_{\rm d}$ [M$_\odot$]} & \colhead{$\chi^2_r$} & & \colhead{$\Delta$BIC}}
\startdata
\\
Dwarfs & &$1.41^{+0.05}_{-0.04}$ & $-0.72^{+0.03}_{-0.04}$ & $9.18^{+0.22}_{-0.26}$ & $6.59^{+0.38}_{-0.38}$  & $4.91$ & & $10.52^{+0.03}_{-0.01}$& $6.25^{+0.57}_{-0.57}$ & $25.5$ & &$-250$\\
\\
\hline
\\
HSB1 & & $1.38^{+0.07}_{-0.06}$ & $-1.07^{+0.11}_{-0.09}$ & $9.68^{+0.31}_{-0.36}$ & $7.38^{+0.79}_{-0.89}$ & $1.37$  & & $11.69^{+0.03}_{-0.03}$ & $6.78^{+0.51}_{-0.51}$ & $4.36$ & &$-50$\\
\\
HSB2 &  & $1.35^{+0.06}_{-0.06}$ & $-1.39^{+0.17}_{-0.17}$ & $9.85^{+0.40}_{-0.36}$ & $8.03^{+1.3}_{-0.67}$ & $0.94$ & & $12.34^{+0.06}_{-0.06}$ & $8.81^{+0.20}_{-0.07}$ & $3.54$ & & $-42$\\
\\
HSB3 & & $1.30^{+0.03}_{-0.04}$ & $-1.11^{+0.21}_{-0.13}$ & $10.70^{+0.11}_{-0.16}$& $9.71^{+0.10}_{-0.03}$ & $1.98$ & & $12.24^{+0.04}_{-0.06}$ & $9.62^{+0.06}_{-0.01}$ & $4.87$ & & $-51$\\
\\
HSB4 & & $1.27^{+0.05}_{-0.05}$ & $-1.17^{+0.27}_{-0.17}$& $10.98^{+0.14}_{-0.18}$& $9.87^{+0.21}_{-0.11}$ & $1.70$ & & $12.40^{+0.03}_{-0.08}$ & $9.88^{+0.09}_{-0.06}$ & $3.53$ & &$-30$\\
\\
HSB5 & & $1.03^{+0.10}_{-0.03}$ & $-0.72^{+0.89}_{-0.89}$ & $12.16^{+0.31}_{-0.20}$ & $9.99^{+0.22}_{+0.18}$ & $17.7$ & & $12.41^{+0.04}_{-0.06}$ & $10.17^{+0.05}_{-0.01}$ & $15.8$ & & $+5$\\
\\
HSB6 & & $1.11^{+0.05}_{-0.04}$ & $-1.59^{+0.37}_{-0.37}$ & $11.74^{+0.27}_{-0.54}$& $10.34^{+0.10}_{-0.09}$ & $0.79$ & & $12.60^{+0.04}_{-0.06}$ & $10.27^{+0.06}_{-0.03}$ & $1.13$ & & $-2$\\
\\
HSB7 & &$1.08^{+0.07}_{-0.07}$ & $-0.68^{+0.78}_{-0.78}$ & $12.28^{+0.37}_{-0.31}$& $10.61^{+0.03}_{+0.04}$ & $0.52$ & & $12.45^{+0.05}_{-0.07}$ & $10.54^{+0.05}_{-0.02}$ & $0.47$ & & $+5$\\
\\
HSB8 & & $1.02^{+0.05}_{-0.05}$ & $-0.64^{+0.86}_{-0.86}$ & $12.16^{+0.15}_{-0.07}$ & $10.82^{+0.05}_{+0.02}$ & $3.74$ & & $12.24^{+0.06}_{-0.09}$ & $10.79^{+0.06}_{-0.04}$ & $3.33$ & &  $+5$\\
\\
HSB9 & & $1.09^{+0.09}_{-0.09}$ & $-0.40^{+0.76}_{-0.76}$ & $12.41^{+0.27}_{-0.37}$ & $10.97^{+0.06}_{-0.04}$ & $1.42$ & & $12.48^{+0.05}_{-0.09}$ & $10.88^{+0.08}_{-0.06}$ & $1.26$ & & $+5$\\
\\
HSB10 & & $1.11^{+0.09}_{-0.09}$ & $-0.29^{+0.78}_{-0.78}$ & $12.45^{+0.03}_{-0.46}$ & $11.19^{+0.06}_{-0.04}$ & $2.25$ & & $12.26^{+0.05}_{-0.08}  $ & $11.11^{+0.07}_{-0.05}$ & $1.99$ & & $+5$\\
\\
HSB11 & & $1.12^{+0.09}_{-0.09}$ & $-0.02^{+0.65}_{-0.65}$ & $12.56^{+0.10}_{-0.34}$ & $11.43^{+0.04}_{-0.03}$ & $0.66$ & & $12.39^{+0.07}_{-0.09}$ & $11.35^{+0.06}_{-0.03}$ & $0.59$ & & $+6$\\
\\
\hline
\\
LSB1 & & $1.36^{+0.11}_{-0.06}$ & $-0.31^{+0.07}_{-0.19}$ & $9.93^{+0.42}_{-0.55}$ & $7.57^{+1.1}_{-0.6}$ & $1.84$ & & $10.46^{+0.04}_{-0.01}$ & $6.67^{+0.66}_{-0.66}$ & $12.3$ & & $-103$\\
\\
LSB2 & & $1.08^{+0.09}_{-0.09}$ & $-1.14^{+0.79}_{-0.79}$ & $10.83^{+0.67}_{-0.03}$ & $8.17^{+0.90}_{-0.15}$ & $0.65$ & & $11.29^{+0.06}_{-0.02}$ & $8.39^{+0.47}_{-0.06}$ & $0.52$ & & $+5$\\
\\
LSB3 & & $1.23^{+0.06}_{-0.10}$& $-0.17^{+0.54}_{-0.49}$ & $11.44^{+0.60}_{-0.69}$ & $9.81^{+0.27}_{-0.03}$ & $0.25$ & & $11.64^{+0.04}_{-0.04}$ & $9.29^{+0.17}_{-0.05}$ & $2.37$ & & $-16$\\
\\
LSB4 & & $1.03^{+0.06}_{-0.06}$ & $-0.80^{+0.84}_{-0.84}$ & $11.39^{+0.18}_{-0.04}$ & $10.35^{+0.13}_{+0.07}$ & $2.42$ & & $11.48^{+0.12}_{-0.03}$ & $10.43^{+0.04}_{-0.02}$ & $1.74$ & & $+4$\\
\\
LSB5 & & $1.05^{+0.09}_{-0.09}$ & $-0.70^{+0.91}_{-0.91}$ & $11.64^{+0.36}_{-0.28}$ & $10.73^{+0.39}_{-0.36}$ & $8.9$ & & $11.99^{+0.02}_{-0.06}$ & $11.04^{+0.08}_{-0.05}$ & $8.8$ & &  $-9$\\
\\
\enddata
\tablecomments{Note that for values $s\lesssim 1.1$ the estimate of $\ell$ is highly uncertain since the fractional model features a rotation velocity approximately independent of such parameter and tends to become very similar to Newtonian gravity.}
\end{deluxetable*}

\clearpage

\clearpage

\clearpage

\clearpage

\clearpage

\clearpage

\clearpage

\clearpage

\clearpage

\clearpage

\end{document}